\documentclass{aa}
\usepackage{graphicx}
\newcommand{\lsim}{\raisebox{-5pt}{$\;\stackrel{\textstyle <}{\sim}\;$}}
\newcommand{\gsim}{\raisebox{-5pt}{$\;\stackrel{\textstyle >}{\sim}\;$}}

\begin{document}
   \title{  Extinction  and metal column density
   \\ of HI regions up to redshift $z \simeq 2$   
}

   \author{Giovanni Vladilo\inst{1}, Miriam Centuri\'on\inst{1}, Sergei A. Levshakov\inst{2}, 
   Celine P\'eroux\inst{3}, \\ Pushpa Khare\inst{4}, Varsha P. Kulkarni\inst{5},
   Donald G. York\inst{6,7}
          }
          
\titlerunning{Extinction and metal column density at high redshift}
\authorrunning{Vladilo et al.}

   \offprints{G. Vladilo}

   \institute{
              Osservatorio Astronomico di Trieste, Istituto Nazionale di Astrofisica,  Trieste, Italy \\
              \email{vladilo@ts.astro.it}
              \and
              Department of Theoretical Astrophysics, Ioffe Physico-Technical Institute,  St. Petersburg, Russia
              \and
              European Southern Observatory, Garching-bei-M\"unchen, Germany 
              \and
              Department of Physics, Utkal University, Bhubaneswar, India
              \and
              Department of Physics and Astronomy, University of South Carolina, Columbia, USA
              \and
              Department of Astronomy and Astrophysics, University of Chicago, Chicago, USA
              \and
              Enrico Fermi Institute, University of Chicago, Chicago, USA         
                }

   \date{Received ...; accepted ...}

 \abstract{ 
We  used the photometric database
of the Sloan Digital Sky Survey (SDSS)
to  estimate the reddening
of  13 SDSS quasars  selected on the basis 
of the presence of  zinc  
absorption lines
  in an intervening Damped Ly $\alpha$ (DLA) system.
In 5 of these quasars the reddening is detected at $\gsim$ 2$\,\sigma$ confidence level 
in two independent color indices  of  the SDSS $ugriz$ photometric system.  A detailed analysis of the data supports an origin of the reddening in the intervening absorbers.  
We used these rare measurements of extinction in DLA systems
to probe the relation between extinction and metal column density
in the interval of absorption redshift $0.7 \lsim z \lsim 2.0$.
We find that the mean  extinction 
 in the $V$ band  per atom of iron  
in the dust is remarkably similar to that found in interstellar clouds of
the Milky Way.  
This result lends support to
previous estimates of the dust obscuration effect in DLA systems
 based on a Milky Way extinction/metal column density relation.  
We propose a simple mechanism, based on dust grain destruction/accretion properties,
which may explain the approximate constancy of the extinction per atom of iron
in the dust.  
   }

   \maketitle
%

\section{Introduction}

Interstellar dust is a pervasive component of galaxies  and
plays a key role in a variety of astrophysical processes affecting, with its presence,
many types of observations of the local
and high-redshift Universe (Spitzer 1978, Draine 2003, Meurer 2004). 
Understanding whether or not the dust at high redshift shares 
similar properties 
with that at low redshift  is central to correctly interpreting the
observations of 
galaxies and quasars  in 
the early Universe. 

In this paper we focus our attention on a particular property
of the dust, namely the extinction per unit column density
of metals. 
The aim of this work
is to estimate this quantity in individual clouds at high-redshift  
and compare it with measurements performed in local clouds. 
The results of this investigation are relevant for 
studies of the extinction generated by quasar absorption line systems
and for casting light on
the physics of dust grains  at different cosmic epochs.


%
From  Milky Way interstellar studies it is well known that the
extinction per  H atom is
roughly constant, with a typical value
$ A_V / N_\mathrm{H}  \approx 5.3 \times 10^{-22}$ mag cm$^{2}$
(Bohlin et al. 1978), where $A_V$ is the extinction in the $V$ band.
The relation is valid over a  wide
interval of column densities, i.e. $5 \times 10^{19} \lsim N_\mathrm{H} \mathrm{ [cm^{-2}]}
\lsim 3\times 10^{21}$.
Observations of translucent clouds (Rachford et al. 2002)
indicate that the same relation holds at high values of extinction
 (up to $A_V \simeq 3$ mag), 
albeit with a modest trend of $ A_V / N_\mathrm{H}$
with $R_V \equiv \nobreak{A_V / (A_B - A_V)}$ (Draine 2003). 
%

The fact that  $A_V$ scales linearly with $N_\mathrm{H}$   
over a wide range of physical conditions probably reflects
an underlying property of dust grains.
If this property is universal, 
we expect to find its signature in high redshift clouds.
To investigate this possibility we must understand
which is the physical relation
underlying the empirical relation $A_V \propto N_\mathrm{H}$. 
Since the extinction is generated by the dust particles, 
 $A_V$  should scale with the column density of atoms in the dust  
rather than with the column density of hydrogen in the gas.
For 
Galactic clouds, it is easy to show that 
$N_\mathrm{H} \propto \widehat{N}_\mathrm{R}$,
where
$\widehat{N}_\mathrm{R}$ is  
the  dust-phase column density of   refractory
elements\footnote{
The symbol $\widehat{N}$ is introduced to indicate
column densities of atoms in the dust phase.
The validity of the relation $N_\mathrm{H} \propto  \widehat{N}_\mathrm{R} $
is the result of the constant (solar) composition of the local
interstellar medium. 
If the composition is constant,
the   gas-phase column density  of volatile elements
scales linearly with the   dust-phase column density  of   refractory elements 
(Vladilo \& P\'eroux 2005). 
Hydrogen is mostly present in the gas phase
and is, in practice,  a volatile element. 
}.
Therefore,
the empirical  Milky Way relation $A_V \propto N_\mathrm{H}$
  might be the result of a more universal
relation $ A_V \propto \widehat{N}_\mathrm{R}$.
 The scientific goal of our investigation is to test  
the existence of this relation 
from a study of high redshift interstellar clouds.
If the relation exists, it should also be able to predict the 
values  of  $A_V / N_\mathrm{H}$
observed in the Magellanic Clouds, which are lower than
in the Milky Way,
even if relatively constant within
each  galaxy (e.g. Gordon et al. 2003).

 To accomplish our goal, we need high redshift
 measurements of extinction and  column densities,
which are rather difficult to obtain. 
 To measure the   column densities one needs a
bright, point-like source located beyond the cloud. Among sources  
at cosmological distances, 
 QSOs are the best targets given the  
very fast decline of Gamma Ray Bursts.
To measure the extinction one needs to estimate which fraction of QSO
light   has been removed from the line of sight 
due to   absorption and scattering processes
within the cloud. This task is particularly difficult owing
to our poor knowledge of the  true spectral energy distribution of 
individual QSOs which, in addition, is   affected by variability.

A way to   overcome these difficulties  
is to  compare composite spectra of quasars with and without foreground 
absorption systems
in a  search for a systematic change of the continuum slope.
After the pioneering work by Pei et al. (1991)
and the lack of     reddening signal
from DLA systems
at redshift $z \sim 2.8$ (Murphy \& Liske 2004),
this approach led to the detection of
reddening   in front of
\ion{Ca}{ii} systems (Wild \& Hewett 2005, Wild et al. 2006) 
and \ion{Mg}{ii} systems (Khare et al. 2005, York et al. 2006)
  at lower redshift.

In the present work we do not follow the same approach, 
since we are interested in  individual absorbers rather than in statistical samples.
In particular, we are interested in studying DLA systems, the QSO absorbers with highest    
   \ion{H}{i} column density,   $N(\ion{H}{i}) > 2 \times 10^{20}$ atoms cm$^{-2}$,
 that  trace the interstellar medium of high redshift galaxies
 (Wolfe et al. 2005).    
Our goal is to estimate the reddening of individual DLA systems
by means of photometric techniques that can be efficiently implemented
in large databases.

Studies of photometric reddenings  of individual quasars with
foreground DLA systems  have started only recently
(Khare et al. 2004, Ellison et al. 2005).
Two of Ellison et al.'s (2005) DLA quasars seem to be significantly reddened,
but in one of these two cases the reddening may well be due to dust
in the quasar host galaxy. 
A preliminary study of the reddening versus gas-phase depletion of the absorber,
based on a very small sample, has been reported 
by Khare et al. (2004).
The behaviour of the rest frame extinction versus metal column density
of the absorber,
the object of the present study, has not been investigated. 
In the previous photometric studies the intrinsic color of the quasars was estimated   using
a   composite spectrum of quasars obtained  from 
the Sloan Digital Sky Survey (SDSS) database (e.g. Richards et al. 2001).
The  error associated with  the composite spectrum
was not propagated into the reddening measurement.

In the present   
 work we measure the reddening  
using the quasar photometric database
of the SDSS (e.g. Schneider et al. 2005), 
paying
special attention to the treatment of the errors.
Instead of  using  a  quasar
composite spectrum, we build up
a distribution of quasar colors 
at each redshift
to estimate both the intrinsic color
(the median of the distribution)
 and its error (the dispersion of the distribution). 
This error is then propagated into the reddening
measurement. 
 At variance with previous work,
we estimate the
reddening of each quasar in at least two different color indices. 
 Using different colors
yields   two advantages. First, we
can  minimize local systematic errors (e.g. contamination
of the quasar continuum in one specific passband of the spectrum). 
Second, as we show below, we can test if the reddening originates
in the intervening system. 

  In the selection of the targets we only consider quasars
with a single DLA system.  In this way 
we can unambiguously investigate
the relation between quasar extinction and column densities
of the absorber. 
In the next section we 
 describe   the    sample, while 
in Sect. 3 we present the reddening measurements
and their conversion to rest-frame extinction.   
In Sec. 4 the  extinction is compared to the  dust-phase column density
of the refractory element iron, $\widehat{N}_\mathrm{Fe}$, 
to test the existence of a general relation $ A_V \propto \widehat{N}_\mathrm{R}$
in high redshift and local interstellar clouds. 
 The results are discussed in Sect. 5    
  and the whole work is summarized in Sect. 6.  
 


\section{ The sample }

In order to generate our sample we followed three criteria. 
The first criterion was driven by the
requirement of estimating the dust-phase column density of iron.
As we shall see below, this can be obtained 
from a simultaneous measurement of the zinc and iron column densities.
We  therefore gathered all the quasar absorption systems with
detections of \ion{Zn}{ii} and \ion{Fe}{ii} lines.

Only absorption systems originating in neutral regions
with $N(\ion{H}{i}) \gsim 2 \times 10^{20}$ atoms cm$^{-2}$,
 i.e. damped Lyman $\alpha$ (DLA) systems, 
were considered.
In these regions, \ion{Zn}{ii} and \ion{Fe}{ii}
are expected to be the dominant ionization stages of zinc and iron 
and the column densities of these metals can be determined without
applying ionization corrections (see Vladilo et al. 2001 and refs. therein).   

In the selection process
we used our previous compilations of DLA systems (Vladilo 2004, Kulkarni et al. 2005)
supplemented by results from recent work (Wang et al. 2004, Akerman et al. 2005,
P\'eroux et al. 2006, Meiring et al. 2006), 
collecting a total of about 70 absorbers. 

The second criterion was to 
 choose, from the resulting list,    
only the DLA systems  with the  quasar  included in the SDSS 
photometric database.  
We used in most cases the Data Release 3
 (DR3)   catalog by Schneider et al. (2005),
supplemented by the Data Release 4 (Adelman-McCarthy et al. 2006) when necessary.
For a bright QSO not included in these catalogs (see Table 1)
we used the list published by Richards et al. (2001).

The homogeneous database of SDSS photometry 
 is an ideal tool for   building up a control sample of quasar colors to be used in
the reddening measurement. 
 Thanks to the large size of the SDSS catalog  
one can easily obtain a control   sample with large numbers
 even in very narrow bins of redshift and magnitude. 
The presence of simultaneous photometry
in the 5 $ugriz$ filters 
allows us to obtain reliable 
reddening measurements in different color indices. 
Estimates of the mean absorber reddening based on photometric measurements
have been shown to be reliable by comparison 
with those based on   spectroscopic measurements 
(York et al. 2006).

The third criterion was to exclude lines of sight
with multiple DLA systems. In fact, for these cases 
the association between quasar reddening and metal column density
would be ambiguous.  
Lines of sight with one neutral region and one (or more) regions of high ionization  
(e.g. \ion{C}{iv}, \ion{Si}{iv})   
were   kept in the sample, assuming the reddening of the ionized region(s) to be negligible compared to that of the neutral region.  
As a result of this     selection,
we obtained the sample of  13 pairs of QSOs/DLA systems shown\footnote
{
The SDSS identifier given in Table 1 is truncated to 4 digits in RA and  DEC
in the rest of the paper.
}
in Table 1.
The large rate of rejection  
relative to the original  \ion{Zn}{ii}  sample    is probably due to 
several factors. One could be
the poor overlap in magnitude space between the quasars of the  \ion{Zn}{ii} sample,
mostly   bright ($m \lsim 18$),
and those of the SDSS catalog, mostly fainter.
Another could be the limited sky coverage of the SDSS. 
Only a very few SDSS quasars with detected Zn were rejected because 
of the presence of multiple DLA systems.

The basic data for the selected absorption systems 
are given in the last 5  columns of   Table 1. 
The original spectroscopic studies of the metal lines  
 were based on observations collected 
with the  ESO VLT telescope (4 cases), the Multiple Mirror Telescope (4 cases),
the Keck telescope (3 cases), the 4-m telescope at the Kitt Peak National Observatory
(1 case) and the {\em HST} (1 case).

For most of the selected systems the DLA nature is confirmed by a direct measurement
of the damped Ly\,$\alpha$ profile. At redshift $z_\mathrm{abs} > 1.9$ the   $N(\ion{H}{i})$
measurements are based on   observations obtained at the   ground-based facilities
mentioned above.
The   Ly\,$\alpha$ measurements at $z_\mathrm{abs} < 1.9$  
are based on UV observations ({\em HST} in 6 cases and {\em IUE} in 1 case).
For two systems at $z_\mathrm{abs} < 1.9$ without   UV observations,
the DLA nature is suggested either by the unusually     strong \ion{Zn}{ii} lines
 (J0121+0027)
 or  by the presence of
 many species of low-ionization, including \ion{Mg}{i} (J2340$-$0053).

As a final check  of the sample of Table 1, we searched for absorption lines 
indicative of potential sources of reddening, even if not classified as DLA systems.  
In this search
we used the low-resolution SDSS spectra of the quasars, plus all the information
presented in previous studies of high resolution spectra. 
A direct search for strong Ly\,$\alpha$ lines at $z < z_\mathrm{qso}$
was not particularly useful since  
for all except one of our targets  
the portion of Ly\,$\alpha$ forest
covered by the SDSS spectrum is too narrow or non-existent. 
However, the wavelength coverage is sufficiently large for detecting \ion{Mg}{ii}
absorptions over a large redshift interval ($0.36 < z_\mathrm{abs} < 2.29$). 
We expect that any intervening \ion{H}{i} cloud with high extinction would produce
a strong \ion{Mg}{ii} resonance doublet. 
In fact, it is now proven observationally that
\ion{Mg}{ii}  absorbers do contribute to the reddening of the quasars (York et al. 2006)
and therefore can be used to trace additional sources of dust.
We therefore searched for narrow, but intense absorption features 
that could be attributed to \ion{Mg}{ii}  systems
at a redshift different from that of the selected DLA system.   
In this search we also checked for 
\ion{Mg}{ii} absorptions superposed on  the \ion{Mg}{ii} quasar emissions,
an indirect signature of dust in  the quasar host galaxy.
In most cases we found no signature of strong \ion{Mg}{ii} lines
or other strong lines from \ion{H}{i} regions,
{\em either from intervening systems or from associated systems (quasar host galaxy)}.

In two quasar spectra, however, we did find evidence for additional absorption
from neutral gas. 
These quasars are listed separately at the end of Table 1 since for these two
lines of sight an additional contribution to the reddening might be present,
in addition to that expected from the DLA system.
Details on the additional absorptions are reported 
in the footnotes to the table. 
With the possible exception of these two cases, 
the DLA systems of Table 1 are very likely to be the major
source of reddening of their respective quasars. 

Given the pre-selection in \ion{Zn}{ii} and \ion{Fe}{ii},  
  the sample of Table 1  does not represent a random sub-set of the total DLA population.
Given the current limitations 
of high resolution spectroscopy at   $\lambda \gsim 800$\,nm,
the pre-selection in \ion{Zn}{ii} precludes   the range of absorption
redshift $z \gsim 3$. 
In spite of these limitations, the list of Table 1 is the best currently available sample for   studying
possible correlations between reddening and metal column densities of individual systems,
considering that reliable \ion{Zn}{ii} and \ion{Fe}{ii} column densities 
are extremely difficult to derive from SDSS data for individual objects.

 \begin{center} 
\begin{table*} 
\scriptsize{
\caption{SDSS quasars with previously detected \ion{Zn}{ii} and \ion{Fe}{ii} absorptions 
in an intervening Damped Ly$\alpha$ system. }
\begin{tabular}{lllccclcl}
\hline \hline 
SDSS & Other name & $z_\mathrm{qso}$ & $g$ & $z_\mathrm{abs}$ & 
$\log N(\ion{H}{i})$ &  Ref.$^a$ & [Zn/H]$^b$ &  Ref.$^c$  \\
 &  & & (mag) &   & (cm$^{-2}$) & &  & \\
 \hline
& & & & & & & \\
J001306.1+000431 & LBQS 0010$-$0012 & 2.165 & 18.65 & 2.025 & $20.80 \pm 0.10$ & LPS03  & $-1.05 \pm 0.10$ & LPS03  \\
J001602.4$-$001225 & Q0013$-$004 & 2.087 & 18.28 & 1.973 & $20.83 \pm 0.05$& PSL02 &$-0.65 \pm 0.06$ & PSL02 \\
J012147.7+00271 &   B0119+0011 & 2.224 &  20.00 & 1.388 & $[21.04]^d$ & W04 & --- & W04,  P05 \\
J093857.0+412821 & Q0935+417 & 1.936 & 16.49 & 1.373 & $20.52 \pm 0.10$& LWT95 &$-0.92 \pm 0.11$ & MLW95 \\
J094835.9+432302 & Q0948+433 & 1.892 & 18.10 & 1.233 & $21.62 \pm 0.06$ & R05 & $-1.12 \pm 0.06$ & R05 \\
J101018.1+000351 & & 1.400 & 18.27 & 1.265 & $21.52 \pm 0.07$ & R06 & $-1.02 \pm 0.09$ & M06 \\   
J110729.0+004811 & & 1.391 & 17.64 & 0.741 & $20.98 \pm 0.15$ & R06 & $-0.60 \pm 0.16$ & K04 \\
J115944.8+011206 & Q1157+014 & 2.000 & 17.59 & 1.944 & $21.80 \pm 0.10$ & WB81 & $-1.36 \pm 0.13$ & PSL00 \\
J123200.0$-$022404 & PKS 1229$-$021& 1.044 & 17.13 & 0.395 & $20.75 \pm 0.07$ & B98 & $-0.47 \pm 0.14$ & B98 \\
J132323.7$-$002155$^e$ & & 1.388 & 18.46 & 0.716 & $20.21 \pm 0.20$ & P06, R06  & $+0.57 \pm 0.21$ & K04, P06 \\ 
J150123.4+001939 & & 1.928 & 18.11 & 1.483 & $20.85 \pm 0.05$ & R06 & $-0.40 \pm 0.07$ & M06 \\   
 \hline
J223408.9+000001$^f$ &LBQS 2231$-$0015  & 3.015 & 17.57 & 2.066 & $20.56 \pm 0.10$& LW94 & $-0.75 \pm 0.10$ &PW99 \\
J234023.6$-$005326$^g$ & & 2.085 & 17.76 & 1.361 & $[20.30]^d$& K04 & --- & K04 \\
%
\hline
\end{tabular}
\scriptsize{  
\\
$^a$ References for \ion{H}{i} column density data. 
\\
$^b$
 We adopt the usual definition [X/Y] $\equiv \log ( N_\mathrm{X}/N_\mathrm{Y}) - \log (\mathrm{X/Y})_{\sun}$.
Throughout this paper we use the meteoritic solar abundances  of      
Anders \& Grevesse (1989) for consistency with most previous work on interstellar depletions. 
\\
$^c$ References for \ion{Zn}{ii} and \ion{Fe}{ii} column density data (see also Table 3).
\\
$^d$ Indirect estimate of $\log N(\ion{H}{i})$ published by the authors
 based on reddening determinations.
\\
$^e$
For the absorber at $z_\mathrm{abs}=0.716$ towards this quasar 
the \ion{H}{i} column density      adopted here  (P\'eroux et al. 2006) 
is lower than that   given by Rao et al. (2006), 
$\log N(\ion{H}{i}) = 20.54^{+0.16}_{-0.15}$.
The adopted value formally lies below the DLA  definition threshold, but 
 we keep this system in the list because
the difference ($\simeq 0.1$ dex) is significant only at $\simeq 0.5 \,\sigma$ level
and because, from  a study of the ionization properties of this absorber (P\'eroux et al. 2006),
we do not find differences relative to the properties typical of DLA systems.
\\
$^f$  
The SDSS  photometric data for this  bright quasar (Foltz et al. 1989) can be found in Richards et al. (2001).
No SDSS spectrum is available to date.  A large number of absorption features is present in the spectrum
published by Lu \& Wolfe (1994); in addition of the DLA system at $z=2.066$, these authors report the presence of a strong system 
with both neutral and ionized gas at $z = 2.6527$. 
\\
$^g$ The SDSS spectrum shows a large number of absorption features,
including a \ion{Mg}{ii} doublet at $z_\mathrm{abs} \simeq 2.05$; in  spectra of higher resolution
Khare et al. (2004) find a system at $z=2.0547$ with a mix
of ionization states, including \ion{C}{i}, a   signature of neutral gas.
}
\\
References. B98: Boiss\'e et al. (1998);
K04: Khare et al. (2004); LW94: Lu \& Wolfe (1994);
LWT95: Lanzetta et al. (1995); LPS03: Ledoux et al. (2003); M06: Meiring et al. 2006; MLW95: Meyer et al. (1995);
PSL00: Petitjean et al. (2002); PSL02: Petitjean et al. (2002); P06: P\'eroux et al. (2006); P05: Prochaska (2005, priv. comm.);
PW99: Prochaska \& Wolfe (1999);
R05: Rao et al. (2005); R06: Rao et al. (2006); W04: Wang et al. (2004); WB81: Wolfe \& Briggs (1981).
}
\end{table*}
\end{center}
\normalsize
 

\section{Reddening measurements} 

For each quasar of Table 1 we
measured the reddening in the color index $(y-x)$  
  from the expression 
\begin{equation}
\Delta(y-x)
 = (y-x) \, - \,  (y-x)_\circ
\label{DeltaDef}
\end{equation}
where $(y-x)$ is the observed color of the quasar 
and
$\nobreak{(y-x)_\circ}$ the  intrinsic color 
in absence of reddening.
All quantities in this definition  
are in the observer's frame.

The color $(y-x)$ was  measured
 using different pairs of  
PSF magnitudes 
$x,y=u,g,r,i,z$   of the SDSS DR3 catalog
preliminarily
corrected  for Galactic extinction
(Schneider et al. 2005).

We only used bandpasses falling in spectral regions where
the continuum of the quasar is  
not contaminated by the Ly\,$\alpha$ forest or 
by the quasar Ly\,$\alpha$ emission. 
In practice, 
taking into account the effective wavelengths and FWHM of the $ugriz$
bandpasses (Fukugita et al. 1996, Schneider et al. 2005),
this criterion precludes the use of the $u$ band
when $z_\mathrm{qso} > 1.67$  
and also of the $g$ band when $z_\mathrm{qso} > 2.25$.
As an additional criterion,
pairs of adjacent $ugriz$ bandpasses  were not considered
since wavelength separation is essential for detecting the reddening 
of the quasar, if present. 
The colors obtained with these criteria are given in Table 2, Col. 3.

To estimate the intrinsic color and its  uncertainty
we first generated a control sample of reference quasars
for each quasar of our list.
Homogeneity in redshift and apparent magnitude were
the criteria adopted for building each control sample. 
The homogeneity in redshift is very important 
given the redshift dependence of the quasar colors 
(Richards et al. 2001). 
The large database of the DR3 catalog 
(46420 objects) allowed us to select 
control samples  with a large number
of reference quasars even using small redshift bins. 
In practice, we adopted
$z_\mathrm{ref} 
\in (\nobreak{z_\mathrm{qso}-0.05},\nobreak{ z_\mathrm{qso}+0.05})$
in all cases but one (see note $a$  of Table 2).

After selecting a sample with a given redshift, we applied the 
criterion of homogeneity in apparent magnitude. 
This is in practice equivalent to a homogeneity in absolute luminosity
since the quasars of each sample are at the same redshift.  
We used the infrared magnitude $z$, 
which is the least affected by extinction. 
The width of the magnitude bin was tuned in such a way
as to obtain control samples of the same size for all the quasars
of our list.
For instance,   to obtain a   control sample of 100 quasars, 
starting from the sample already binned in redshift,
the typical width of the magnitude bin  was of   $\approx 2$ magnitudes.
Changes of the intrinsic slopes of the quasar continua over
an interval of  $\approx 2$ in absolute magnitude  are expected to be very modest
 (e.g. Yip et al. 2004).

\begin{center} 
\begin{table*} 
\scriptsize{
\caption{Colors and reddenings of the quasars of Table 1.}     
\begin{tabular}{|ccr|rrr|r|c|}
\hline \hline 
SDSS  & Color & Quasar color &  & Median~ &               & PR$^d$       & Quasar reddening   \\
$z_\mathrm{qso}$ & index & $(y-x)~~$  &  & $(y-x)_\circ$ & &  (\%)  & $\Delta(y-x)$  \\
& & & S300$^a$ & S100$^b$ & Clean$^c$& & \\
%
 \hline
%
J1232$-$0224 
& $(u\!-\!z)$ & $0.515\pm0.023$&$ 0.323$ & $ 0.324 $&$ 0.325$ & 6&$ +0.19^{+0.17}_{-0.34}$ \\
1.044
& $(u\!-\!i)$ & $0.551\pm0.020$&$ 0.322$ &$ 0.322 $&$ 0.323$ & &$ +0.23^{+0.15}_{-0.29}$\\ 
& $(g\!-\!z)$ & $ 0.238 \pm 0.029$ & $ 0.173 $&$ 0.177 $ &$ 0.178  $ & &$ +0.06^{+0.12}_{-0.18}$ \\
& $(u\!-\!r)$ & $0.572\pm0.020$ & $ 0.365 $&$ 0.362 $&$ 0.361 $ & &$ +0.21^{+0.12}_{-0.24}$\\ 
& $(g\!-\!i)$ & $0.275\pm0.027$  & $ 0.176 $ &$ 0.180 $&$ 0.180 $ & &$+0.09^{+0.11}_{-0.11}$ \\    
& $(r\!-\!z)$ & $-0.058\pm0.020$ &$-0.038 $ &$-0.016 $&$-0.018 $ & &$-0.04^{+0.13}_{-0.12}$ \\                                                          
\hline
 J1323-0021 
& $(u\!-\!z)$ & $1.322\pm0.033$ &$0.361$ &$ 0.394$& $ 0.387$ & 6&$+0.94^{+0.19}_{-0.30}$ \\ 
1.388
& $(u\!-\!i)$ & $1.249 \pm  0.033$ &$ 0.385$ &$ 0.400$&$ 0.390$ & &$+0.86^{+0.17}_{-0.26}$  \\
& $(g\!-\!z)$ & $0.827\pm0.033$ & $ 0.255$&$ 0.289$&$ 0.293$ & &$+0.53^{+0.18}_{-0.18}$  \\
& $(u\!-\!r)$ & $1.037\pm0.028$ &$ 0.315$ &$ 0.325$&$ 0.320$ & &$+0.72^{+0.13}_{-0.22}$  \\
& $(g\!-\!i)$ & $0.754\pm0.033$ &$ 0.288$ &$ 0.294$&$ 0.293$ & &$+0.46^{+0.11}_{-0.13}$ \\ 
&  $(r\!-\!z)$ & $0.285\pm0.026$ & $ 0.033$&$ 0.059$&$ 0.055$ & &$+0.23^{+0.10}_{-0.09}$ \\ 
\hline
J1107+0048   
& $(u\!-\!z)$ & $0.459\pm0.029$ & $ 0.351 $&$ 0.364$&$ 0.335$ & 11&$+0.12^{+0.17}_{-0.29}$ \\
1.391
& $(u\!-\!i)$ & $ 0.495 \pm 0.024$ & $ 0.383$&  $ 0.401$&$ 0.377$& & $+0.12^{+0.16}_{-0.23}$ \\
& $(g\!-\!z)$ & $0.342\pm0.027$ & $ 0.256$& $ 0.261$&$ 0.238$ & &$+0.10^{+0.13}_{-0.20}$ \\
& $(u\!-\!r)$ & $0.413\pm0.026$ & $ 0.314$&$ 0.324$&$ 0.305$ & &$+0.11^{+0.12}_{-0.22}$ \\ 
& $(g\!-\!i)$ & $0.378\pm0.022$ &$ 0.290$ &$ 0.281$&$ 0.270$ & &$+0.11^{+0.11}_{-0.15}$  \\                               
& $(r\!-\!z)$ & $0.046\pm0.026$ &$ 0.037$ &$ 0.026$&$ 0.018$ & &$+0.03^{+0.08}_{-0.11}$ \\
\hline
J1010+0003 
& $(u\!-\!z)$ & $ 0.500 \pm 0.037 $ & $0.337 $&  $0.388$&$0.374 $& 5 & $+0.13^{+0.23}_{-0.32}$ \\
1.400
& $(u\!-\!i)$ & $ 0.520 \pm 0.027 $ & $0.356 $&  $0.393$&$0.380 $& & $+0.14^{+0.16}_{-0.31}$ \\
& $(g\!-\!z)$ &$ 0.159 \pm 0.032 $ & $ 0.244 $&  $ 0.253$&$0.250 $& & $-0.09^{+0.13}_{-0.20}$ \\
& $(u\!-\!r)$ &$ 0.486 \pm 0.025 $ & $ 0.294$&  $0.315 $&$0.311 $& & $+0.18^{+0.14}_{-0.28}$ \\
& $(g\!-\!i)$ &$ 0.180 \pm 0.019 $ & $ 0.279$&  $0.289$&$0.283 $& & $-0.10^{+0.11}_{-0.14}$ \\
& $(r\!-\!z)$ &$ 0.014 \pm 0.032 $ & $ 0.039 $&  $0.039$&$0.032 $& & $-0.02^{+0.08}_{-0.13}$ \\
\hline
J0948+4323
& $(g\!-\!z)$ & $ 0.289 \pm 0.057$ & $ 0.365 $ & $ 0.374 $ & 0.339  & 29 & $-0.05^{+0.16}_{-0.15}$ \\
1.892
& $(g\!-\!i)$ & $ 0.397 \pm 0.054$ & $ 0.303 $ & $ 0.302 $ & 0.278 & & $ +0.12^{+0.14}_{-0.13}$ \\
& $(r\!-\!z)$ & $ 0.369 \pm 0.030$ & $ 0.312 $ & $ 0.304 $ & 0.284 & & $ +0.09^{+0.10}_{-0.10}$ \\
\hline
J1501+0019 
& $(g\!-\!z)$ & $ 0.246 \pm 0.024 $ & $0.380 $&  $ 0.375 $&$ 0.344$&27 & $-0.10^{+0.14}_{-0.11}$ \\
1.928 
& $(g\!-\!i)$ & $ 0.215 \pm 0.024 $ & $ 0.274$&  $0.274 $&$0.251$& & $-0.04^{+0.11}_{-0.13}$ \\
& $(r\!-\!z)$ & $ 0.260 \pm 0.026 $ & $ 0.316$&  $ 0.311$&$0.298 $& & $-0.04^{+0.10}_{-0.09}$ \\
\hline
J0938+4128   
&  $(g\!-\!z)$& $0.358\pm0.049$ &$0.386 $ &$0.427 $&$0.370 $&27 &$-0.01^{+0.13}_{-0.15}$  \\
1.936
&  $(g\!-\!i)$ & $0.281\pm0.075$ & $0.274$&$0.298$&$0.270$& &$+0.01^{+0.11}_{-0.13}$  \\
&  $(r\!-\!z)$ & $0.292\pm0.049$ & $0.315$&$0.317$&$0.300$& &$-0.01^{+0.09}_{-0.10}$   \\
\hline
J1159+0112   
&  $(g\!-\!z)$ & $0.789\pm0.023$& $ 0.411$ &$ 0.432$&$ 0.422$ & 20&$+0.37^{+0.13}_{-0.14}$  \\
2.000
&  $(g\!-\!i)$ & $0.545\pm0.023$ &$ 0.258$ &$ 0.273$&$ 0.263$ & &$+0.28^{+0.10}_{-0.13}$ \\ 
& $(r\!-\!z)$ & $0.500\pm0.021$ & $ 0.330$&$ 0.338$&$ 0.337$ & &$+0.16^{+0.12}_{-0.10}$ \\
\hline
J2340$-$0053 
& $(g\!-\!z)$& $0.778\pm0.025$ &$0.421$ &$0.441$&$0.420$& 30 &$ +0.36_{-0.18}^{+0.14}$ \\
2.085
& $(g\!-\!i)$ & $0.536\pm0.021$ &$0.231$ &$0.262$&$0.232$& &$ +0.30_{-0.17}^{+0.10}$ \\ 
& $(r\!-\!z)$ & $0.540\pm0.028$ &$0.320$ &$0.338$&$0.326$& &$ +0.21_{-0.15}^{+0.13}$ \\
\hline
J0016$-$0012  
& $(g\!-\!z)$ & $0.797\pm0.031$&$0.415$  &$0.426$&$0.397$&23 &$+0.40^{+0.13}_{-0.18}$   \\
2.087
&  $(g\!-\!i)$  & $0.493\pm0.028$ &$0.230$ &$0.237$&$0.209$& &$+0.28^{+0.10}_{-0.13}$   \\ 
& $(r\!-\!z)$  & $0.576\pm0.025$ &$0.318$ &$0.337$&$0.323$& &$+0.25^{+0.14}_{-0.16}$ \\
\hline
J0013+0004 
& $(g\!-\!z)$ & $0.305\pm0.044$  & $0.399$ & $0.393$ & $0.362$ &30&$-0.06^{+0.15}_{-0.10}$   \\
2.165
& $(g\!-\!i)$ & $0.116\pm0.039$  & $0.193$&$0.187$&$0.175$& &$-0.06^{+0.12}_{-0.09}$   \\
& $(r\!-\!z)$ & $0.189\pm0.043$  & $0.298$ &$0.291$&$0.283$& &$-0.09^{+0.12}_{-0.11}$  \\ 
\hline
J0121+0027  
&  $(g\!-\!z)$  & $1.218 \pm  0.047 $ & $0.349$ &$0.371$&$0.350$&21&$+0.87_{-0.14}^{+0.14}$     \\
2.224
&  $(g\!-\!i)$  & $0.912 \pm  0.048$ &$0.143$&$0.156$&$0.128$& &$+0.78_{-0.16}^{+0.12}$   \\
&  $(r\!-\!z)$  & $0.644 \pm  0.047$ &$0.297$ &$0.297$&$0.290$&&$+0.35_{-0.10}^{+0.12}$     \\
\hline
J2234+0000   
&  $(r\!-\!z)$ & $0.250\pm0.016$& $0.142$& $0.149$ &$ 0.141$ &37& $+0.11^{+0.13}_{-0.14}$  \\
3.015
 & &  & & & &  & \\
\hline
\end{tabular}
\scriptsize{  
\\
$^a$ 
Median color of the
control sample of the 300 quasars with closest value of $z$ magnitude.\\
 Half width of redshift bin $\delta z  = 0.05$ except for J2234+0000
($\delta z = 0.15$). 
\\
$^b$ 
Median color of the
control sample of the 100 quasars with closest value of $z$ magnitude. \\
~~Half width of redshift bin $\delta z = 0.05$ in all cases. 
\\
$^c$
Median color of the 'Clean'
sub-sample of S100 (the quasars without absorption features in their SDSS spectra).  
\\
$^d$
Percent of quasars rejected from the control sample S100 on the basis of the presence of absorption lines in their spectra.
}
}
\end{table*}
\end{center}
\normalsize

For each control sample and color index of interest
we then derived the distribution of colors
corrected for Galactic extinction.  
The median of the color distribution was adopted as an estimator
of the intrinsic color $(y-x)_\circ$.
Compared to the mean, the median offers the advantage
of being very stable for different choices of the control sample,
even when some quasars with anomalous colors happen to be included
in the sample. 

We performed several tests to assess the robustness of
the median to sampling errors. The results of
one of these tests is shown in Cols. 4 and 5 of Table 2,
where we compare the medians of the control samples
labelled 'S300' and 'S100',
with 300 and 100 reference quasars respectively.
One can see that the differences in the medians are in most cases 
$\lsim 0.01$ magnitudes. Other tests performed using
sub-samples of S300 and S100 with statistically significant
number of objects
yield the same indication for the magnitude of the sampling error.

Since a fraction of the quasars of the control sample might
be affected by reddening, we made a visual inspection
of the SDSS quasar spectra in the 'S100' samples  to search for
signatures of intervening absorption systems which could, in principle,
redden the reference quasars. 
In most quasars of our list
we could not search for intervening Damped Lyman $\alpha$ absorptions since
the Lyman $\alpha$ forest lies out of the observable wavelength range.
We therefore searched for strong, narrow absorption lines
redwards of the forest,
tentatively identified as 
species of low
ionization that might arise in an intervening \ion{H}{i} region. 
Quasars showing absorption lines with residual intensity $\lsim 50\%$
redwards of the quasar Lyman\,$\alpha$ emission
were rejected from each S100 sample. 
Also Broad Absorption Line (BAL) quasars were rejected.
In this way we generated a new control sample,
labelled 'Clean' in Table 2.
The fact that the percentage of rejection tends to increase with the quasar redshift  
(see column labelled 'PR') argues in favour of an origin of
the rejected absorptions in intervening regions.
When the percentage of rejection is significant, one can see that
the median color of the parent sample S100 is systematically
redder than the corresponding median of the
 'Clean' sub-sample.  Even if the effect is modest, 
and often of the same order of the sampling errors,
this result indicates  that the intervening absorptions do contribute to the
reddening of the quasars. Also the dispersion of the color
distribution shows a little, but systematic change, being smaller
after the rejection process. 
We therefore adopted the  color distribution of the 'Clean'  control sample
for our measurements.

The intrinsic color $(y-x)_\circ$ and its error were estimated
from the median and the dispersion of the 'Clean' distribution, respectively.
Even though the dispersion around the median is commonly estimated using the
interquartile range, i.e. the interval between the 25th and the 75th percentiles,
we preferred to adopt the range between the 16th and the 84th percentiles, which is
more conservative and consistent with the range bracketed by $\pm 1$ standard deviations in a normal distribution.
In fact, by definition, the adopted interval brackets 68\% of the area of the
distribution around its central value, 
in the same way as the range $\pm 1 \sigma$ does for a normal distribution.  

The resulting reddening  
is given in the last column of Table 2 for each quasar and color index. 
The  adopted uncertainty of the reddening measurement
was obtained by propagating
the error of the quasar color (column 3 of the table)
and the dispersion of  the color distribution explained above. 
One can see from Table 2 that
quasar reddening is detected at the $\sim 3 \sigma$ level in two objects 
and at the $\sim 2 \sigma$ level in other three objects.

A test of the accuracy of 
the zero point of our measurements is offered
by the quasar J0013+0004, whose foreground DLA system
is   dust-free on the basis of its solar [Zn/Fe] ratio
(see Table 4 and Section 4). 
For this quasar we find zero reddening  within the 
errors in all the color indices considered, as expected for  a dust-free system.


   \begin{figure}
   \centering
 \includegraphics[width=8cm,angle=0]{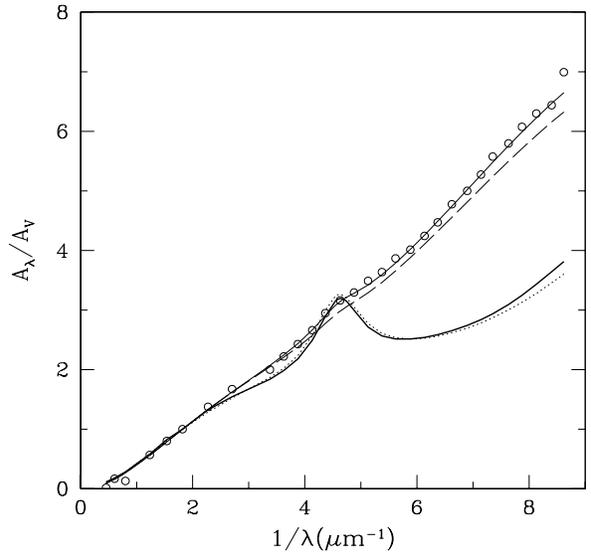}  
  \caption{Comparison of different types of extinction curves normalized to the $V$ band, 
  $\nobreak{\xi(\lambda) = A_\lambda/A_V}$. 
  MW curves: CCM model adopted in this work (thick line) versus P92 model (dotted line).
  SMC curves: modified P92 model adopted here (thin line) to fit the
 G03 data (circles) versus original P92 model (dashed line). See Section 3.1.
 }
              \label{ExtinctionCurves}%
    \end{figure}
%

   \begin{figure*}
   \centering 
  \includegraphics[width=5.8cm,angle=0]{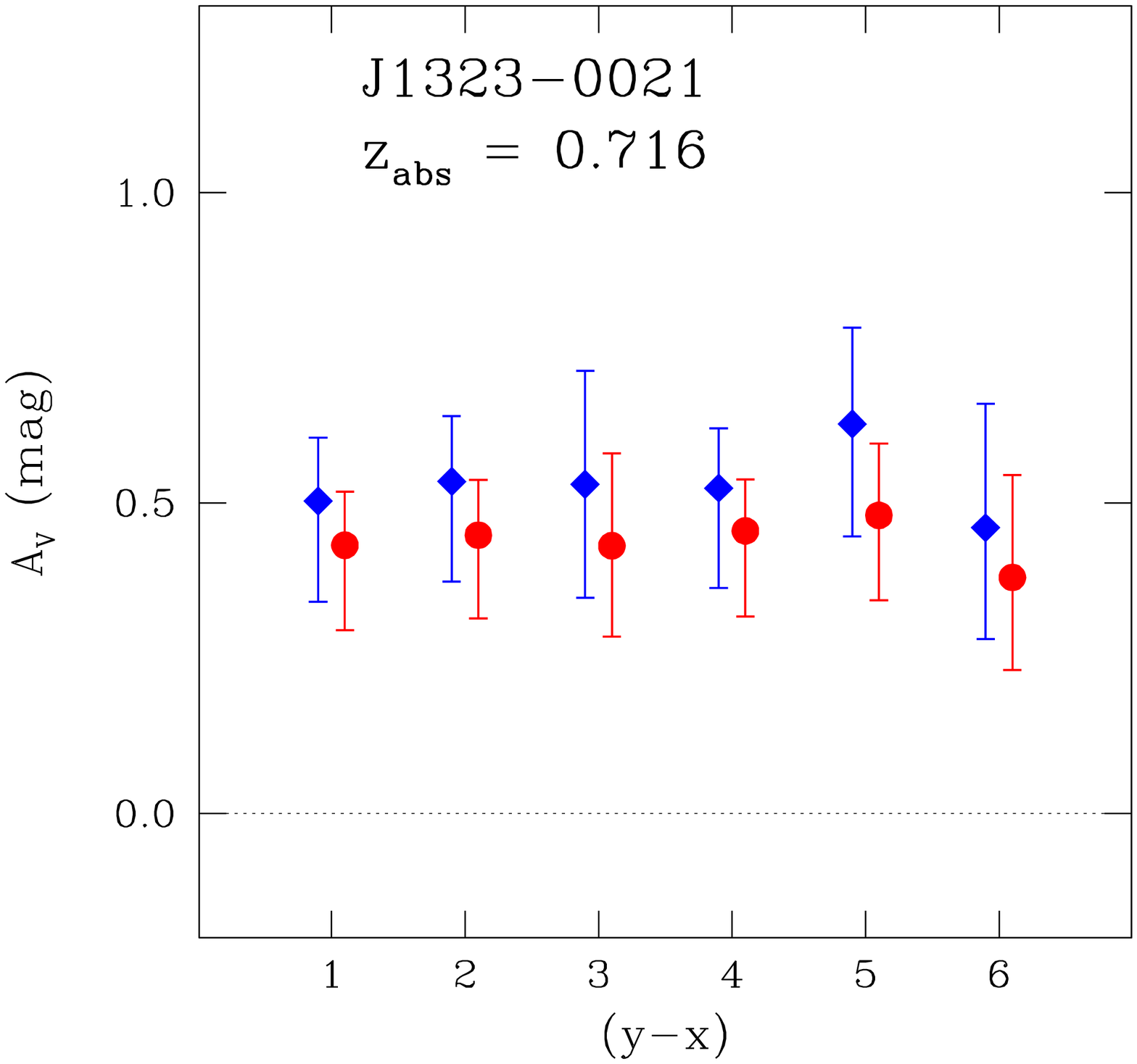}  
  \hskip 0.2 cm
 \includegraphics[width=5.8cm,angle=0]{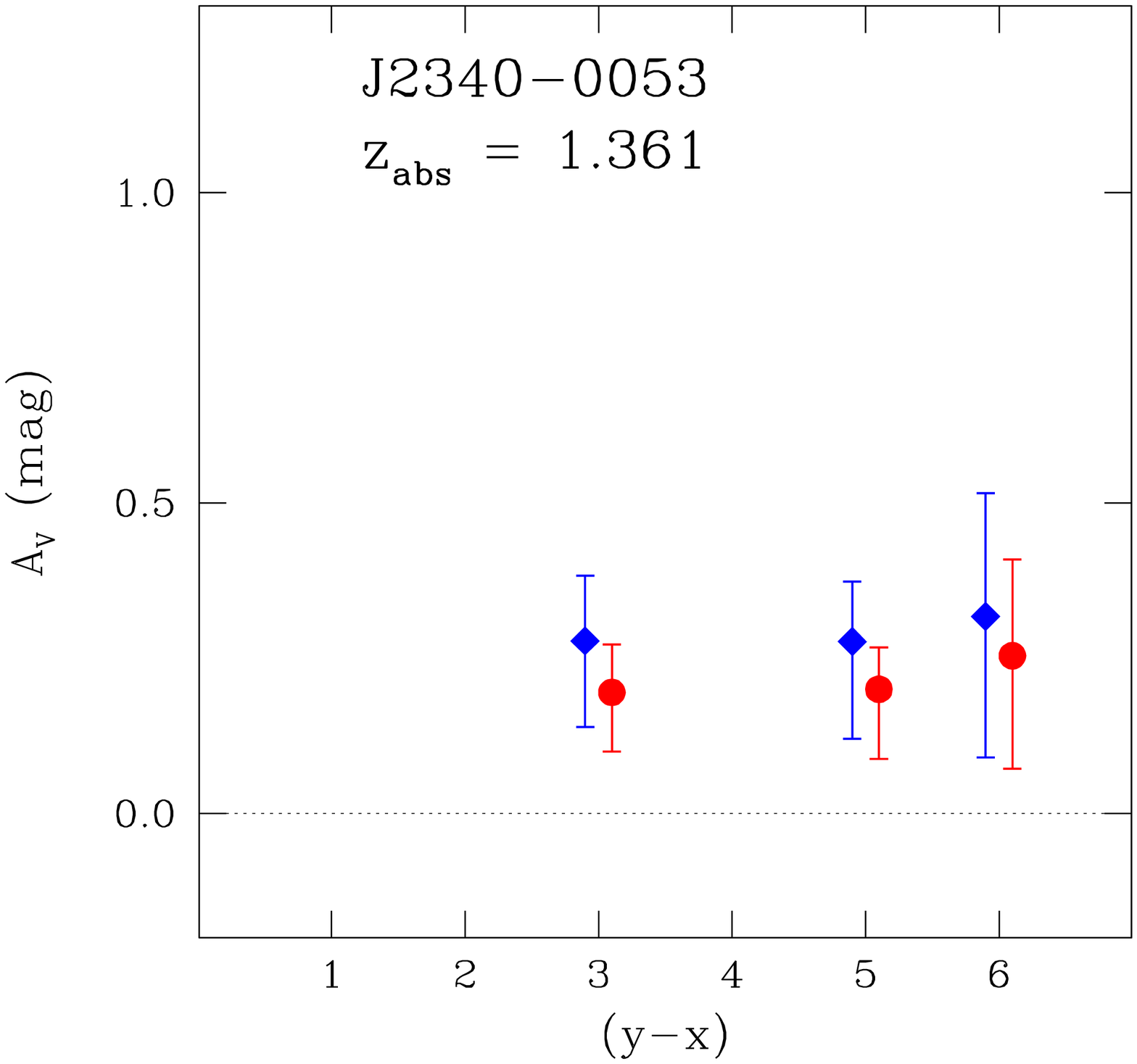} 
  \hskip 0.2 cm
 \includegraphics[width=5.8cm,angle=0]{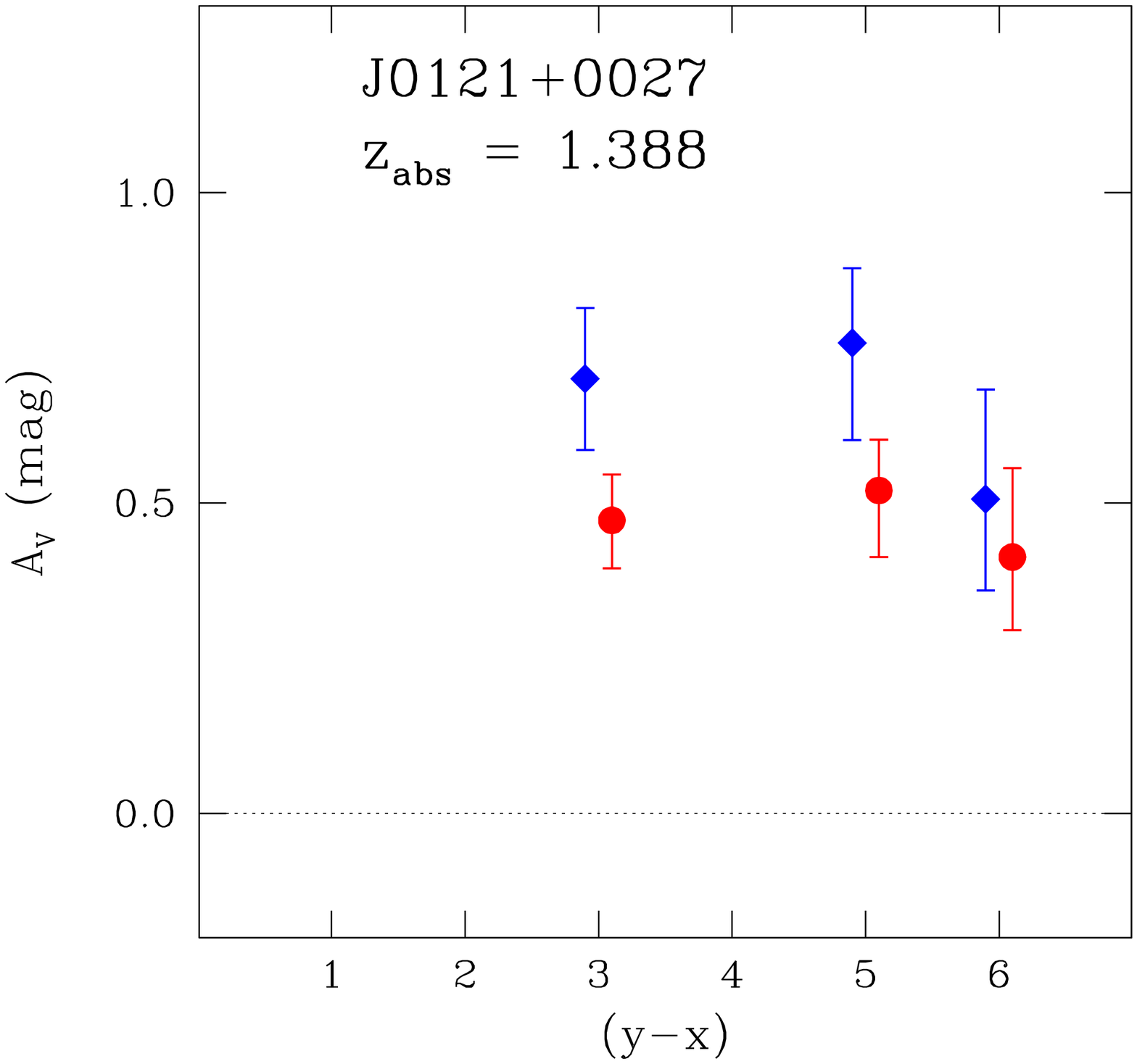} 
  \hskip 5 cm
  \includegraphics[width=5.8cm,angle=0]{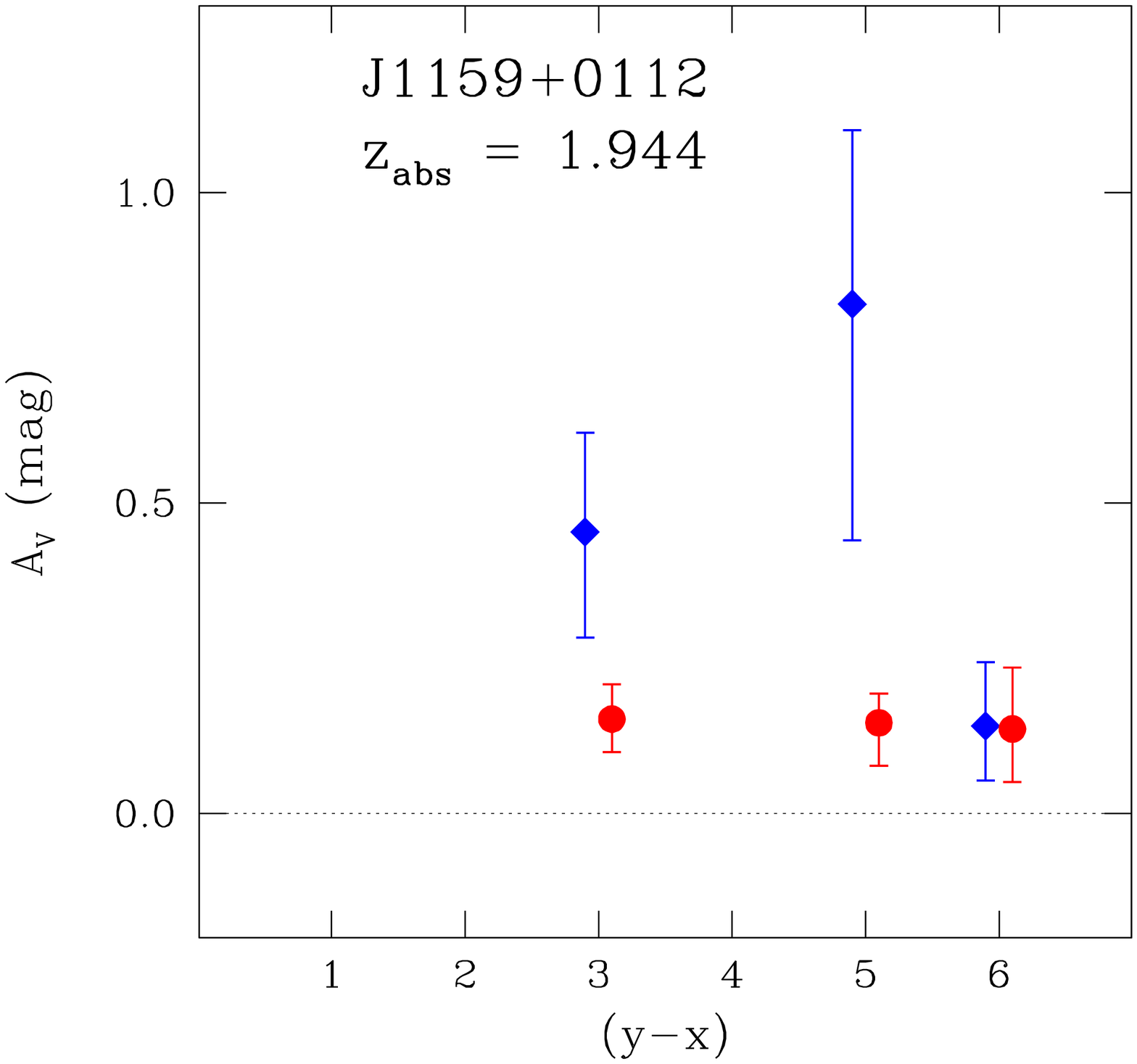}  
 \hskip 0.2 cm
  \includegraphics[width=5.8cm,angle=0]{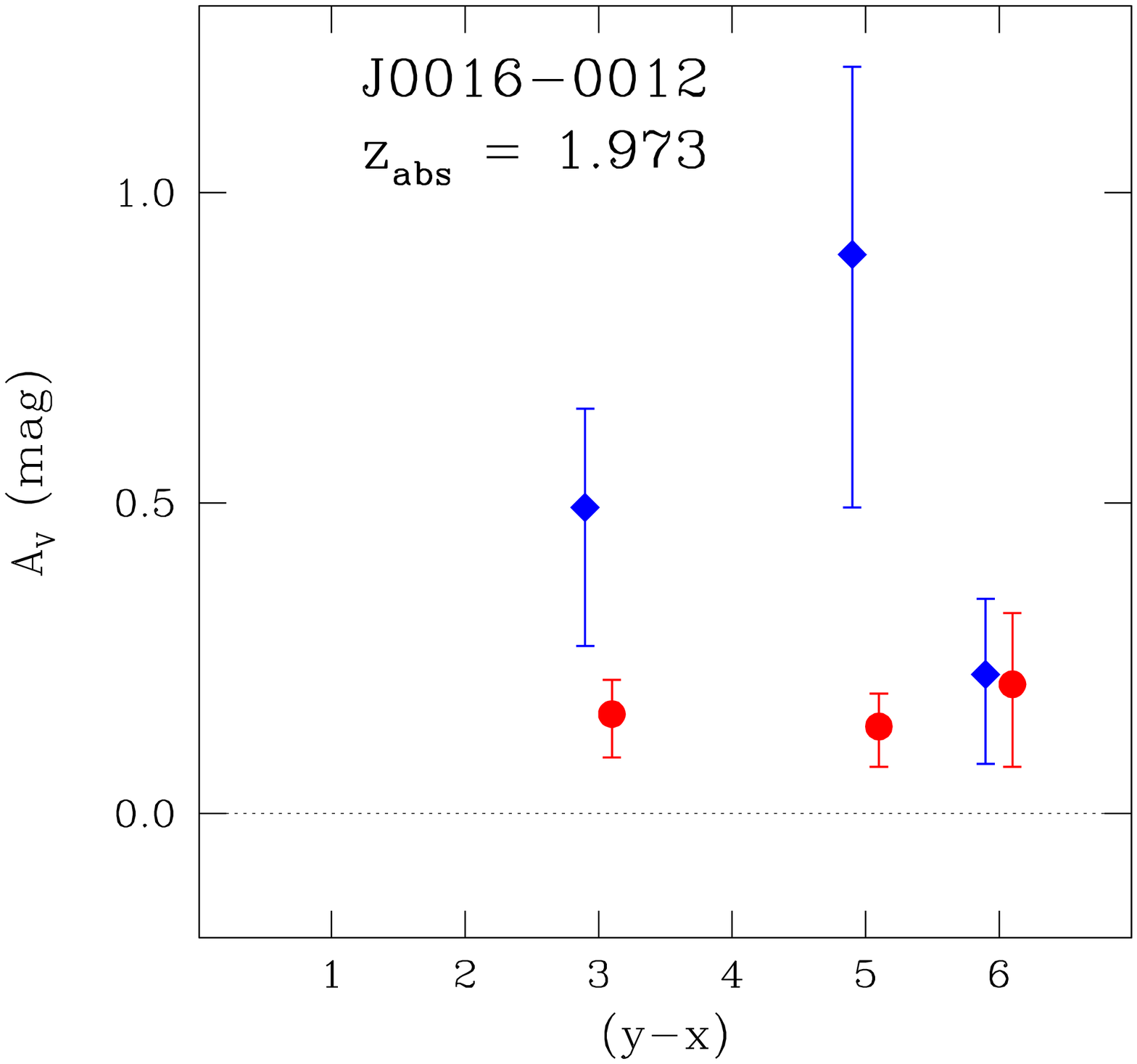}   
  \caption{ 
 Extinction $A_V$ of the reddened quasars of Table 2  estimated in the rest frame of the intervening DLA system  and
 plotted versus color index for different extinction curves.   
 Color index coding: $1=(u\!-\!z)$, $2=(u\!-\!i)$, $3=(g\!-\!z)$, $4=(u\!-\!r)$, $5=(g\!-\!i)$; $6=(r\!-\!z)$.
 Circles:   SMC extinction curve.
 Diamonds:   MW extinction curve.  See Section 3.1.  }
              \label{AVtest}%
    \end{figure*}

\subsection{Conversion to rest-frame extinction}

The quasar reddening measured
in the observer's frame, $\Delta(y-x)$, can be converted
to the $V$ band extinction in the rest-frame of the absorber, $A_V$,
if one knows the normalized extinction curve of the absorber,
$\nobreak{\xi(\lambda) \equiv A_\lambda/A_V}$, and its redshift, $z_\mathrm{abs}$.
 In the Appendix A we show how to perform  this conversion
   taking into account the slope of the quasar continuum.
 If the bandpasses  are sufficiently narrow,
 the variation of the   continuum accross the bandpass can be neglected.
 In this case, 
we obtain the simple relation  useful for the purpose of discussion,
\begin{equation} 
A_V \simeq
{ \Delta (y-x) \over 
\xi({\lambda_{y_m}    \over  1 \! + \!z_\mathrm{abs}}) -  \xi({\lambda_{x_m} \over  1 \!+\!z_\mathrm{abs}})  } ~,
\label{AVsimple}
\end{equation} 
where $\lambda_{x_m}$ and $\lambda_{y_m}$ are the wavelengths of maximum response
of the bandpasses $x$ and $y$. 
The values of $A_V$ presented in our work, however, were derived
from the more general relation (\ref{NumSol}) 
which is also valid for wide bandpasses.  
The response curves of the SDSS $ugriz$ bandpasses were taken from
Fukugita et al. (1996 and priv. comm.). We adopted a power law with spectral index
$\alpha_\nu=-0.5$ for the quasar continuum.

In  our computations we used
the two extinction curves most commonly
adopted in the literature, i.e.
the average curve of the Milky Way,
characterized by an extinction bump at 217.5\,nm
 and that of the Small Magellanic Cloud (SMC),
 characterized by a fast UV rise and without the bump (see e.g. Draine 2003). 
For the first one we adopted the model of
Cardelli et al. (1988; CCM) with $R_V=3.1$.
For the SMC curve we fitted the model of  Pei (1992; hereafter P92) to
the average SMC bar data of Gordon et al. (2003; hereafter G03).  
As we show in Fig. \ref{ExtinctionCurves}, the adopted MW and SMC curves
    are very similar to the models of P92 often used in the literature.

From previous studies we expect the incidence of MW curves to be
very low, compared to that of SMC curves, among quasar absorbers 
(Wild \& Hewett 2005; York et al. 2006 and refs. therein).
However, in a few cases evidence for MW-type curves  
has been reported,
with one clear detection of the bump in a DLA system at $z_\mathrm{abs}=0.524$
observed with the {\em HST} (Junkkarinen et al. 2004), and three
possible identifications in absorption systems
at $z_\mathrm{abs} \simeq 1.4$ - $1.5$ observed in SDSS quasar spectra
 (Wang et al. 2004).

In principle any type of extinction curve can be used in the conversion
from reddening to rest-frame extinction. 
In practice, however, the results obtained from
flat extinction curves, such as the MW curves with $R_V \gsim 5$, 
are unreliable.
This can be understood from the simple relation (\ref{AVsimple}):
if the extinction curve is exactly flat the measured 
$\nobreak{\Delta (y-x)}$ is divided by zero
in the conversion.

\subsection{Comparison of results obtained from different color indices}
 
The rest frame extinction  $A_V$
 is an intrinsic property of the absorber 
and   must be independent of the bandpasses $x$ and $y$
used in the measurement of the reddening. 
Therefore, by comparing values of
$A_V$ obtained from different measurements of $\Delta (y-x)$
in the same quasar
we can test the hypothesis that the measured reddening originates  
in the intervening system.
%
If the hypothesis is correct, we must obtain
a constant $A_V$  from the different values of $\Delta (y-x)$
converted using the redshift $z_\mathrm{abs}$ and the extinction curve $\xi(\lambda)$
of the absorber. 

The result of this test  
depends on the adopted extinction curve. 
However, all known extinction curves share a very similar slope
in the spectral region redwards of the $217.5$\,nm bump
(see Fig. \ref{ExtinctionCurves}).
Therefore, when $z_\mathrm{abs}$ is sufficiently low
for  all the bandpasses   to fall
redwards of the redshifted   bump 
the test is, in practice, independent of the adopted curve. 

In Fig. \ref{AVtest} we plot $A_V$ versus color index
 for the average MW and SMC extinction curves.  
Only   quasars for which the reddening was detected
in at least three colors are shown in the figure.  
One can see in Fig. \ref{AVtest} that in each quasar
$A_V$ is constant within the errors
for at least one of the two extinction curves.  
The test is particularly stringent for  J1323$-$0021,
for which we have at our disposal 6 different measurements.
 These results are consistent with the hypothesis that the reddening 
originates in  the intervening DLA systems.

As expected, the test is unable to discriminate between different types
of extinction curves for the absorbers at lower redshift.
However, when $z_\mathrm{abs}$ is sufficiently high to sample the far UV region
of the extinction curve, the test can also be used to
discriminate between different extinction curves. 
In fact, for the two absorbers at $z_\mathrm{abs} \sim 2$ 
 towards J1159+0112 and J0016-0012
the test  favours an extinction curve of SMC type
which, at variance with the MW curve, yields a constant $A_V$
from the different measurements of $\Delta(y-x)$
(bottom panels in the figure).

In principle, the comparison of the results obtained from different color indices
could also be used to test whether  the 
absorber  originates   at the redshift of the quasar  or not.
We have performed this test for the cases in which $z_\mathrm{qso}$ 
is significantly larger than $z_\mathrm{abs}$, namely
J1323$-$0021, J2340$-$0053 and J0121+0027. 
In all these cases we find that a MW-type extinction curve does not pass the test 
($A_V$ shows a large scatter in different color indices), 
but an SMC extinction curve is still a viable possibility 
($A_V$ is approximately constant). 
We conclude that this test, taken alone, 
is unable to rule out the possibility 
that the reddening originates at the redshift of the quasar. 
This in turn implies that the reddening detection, even if confirmed in different
color indices, must be accompanied by a spectroscopic analysis of the absorption lines,
in order to establish the location along the line of sight
of the absorber responsible for the reddening. 
This conclusion is reinforced by the analysis 
of the quasars without absorption lines in our control samples:
by choosing very reddened quasars in these samples one may obtain
an approximately constant $A_V$   adopting an arbitrary value
of $z_\mathrm{abs} < z_\mathrm{qso}$ in the conversion from reddening to extinction.

   \begin{figure*}
   \centering 
  \includegraphics[width=5.8cm,angle=0]{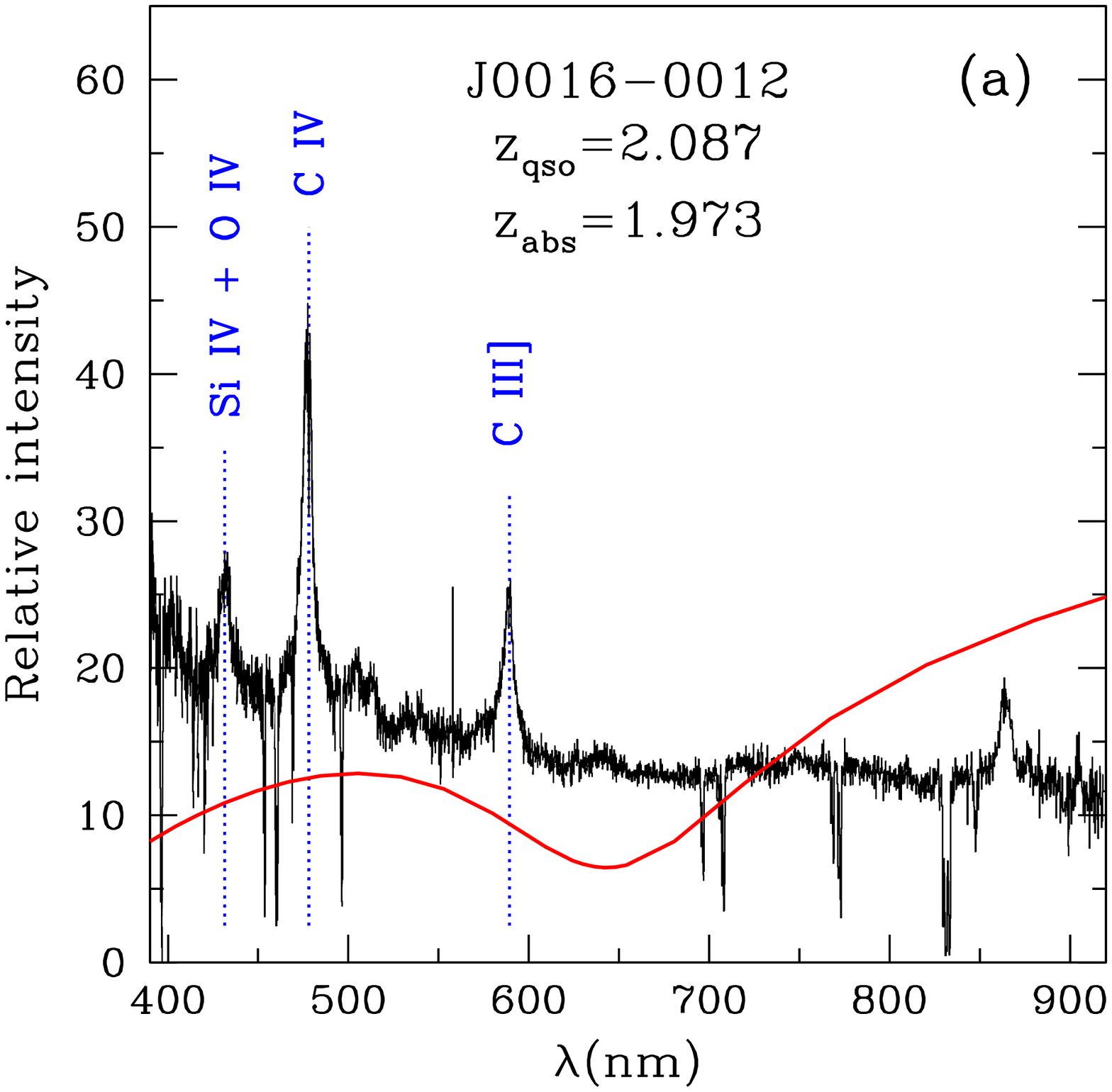}  
  \hskip 0.2 cm
 \includegraphics[width=5.8cm,angle=0]{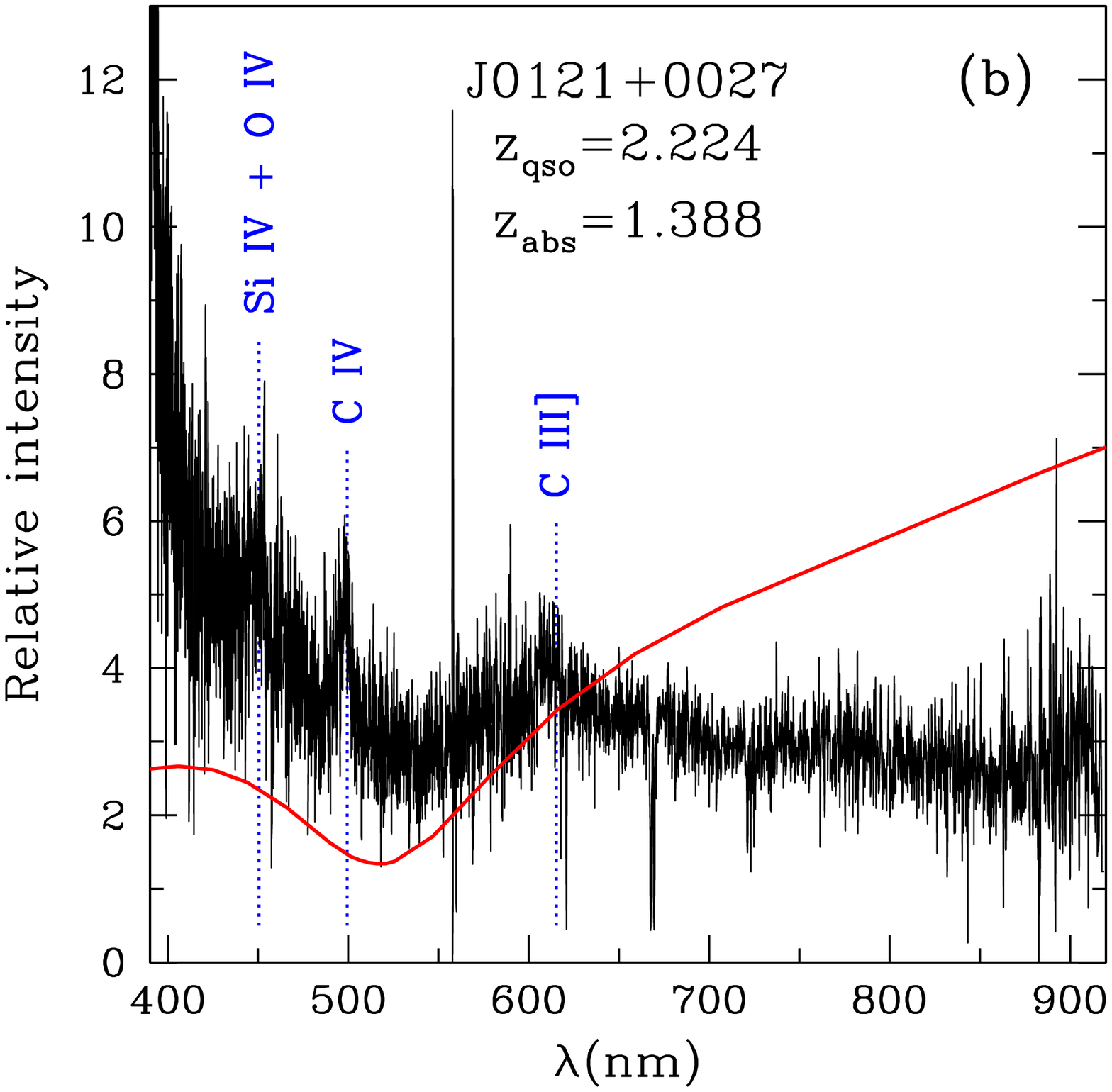} 
 \caption{SDSS spectra of  two
 quasars with a foreground DLA system at redshift $z_\mathrm{abs}$. Smooth lines:    
       MW-type extinction curve ($R_V=3.1$) plotted
 in a scale $\nobreak{10^{-\xi[\lambda/(1+z_\mathrm{abs})]} } $. See Section 3.3. 
  The predicted positions of some emission lines typical of quasar spectra are indicated.
 Intensities are in arbitrary units.
 }
   \label{bump}%
    \end{figure*}

 \subsection{Extinction curve and average $A_V$}

In order to obtain a value of $A_V$ 
for each absorption system of our sample
we adopted one of the two types of extinction curve  
(SMC or MW)
and then averaged the extinction obtained from different color indices.
In practice, we averaged the independent
measurements of $A_V$ obtained from two colors
without common passbands.
Local systematic errors,
such as the contamination of the quasar continuum by emission lines,
are minimized in this way. 
In most cases, we averaged the measurements obtained from the
$(g-i)$ and $(r-z)$ colors. 
For the quasars  with $u$ band
not contaminated by Ly\,$\alpha$ emission/forest  ($z_\mathrm{qso} < 1.67$)
we used the $(u-i)$ and $(g-z)$ colors, which have a larger
spectral leverage for the detection of the reddening.
For J2234+0000 ($z_\mathrm{qso} = 3.015$) we only 
had one uncontaminated color index, i.e. the $(r-z)$.

In order
to choose the extinction curve  we analysed the SDSS spectra of the quasars    
of our sample in a search for the redshifted 217.5\,nm extinction bump.
Since this feature 
is present in the MW curve but absent in the SMC curve, it
can be used to discriminate between the two types of extinction.
Within the limits of the wavelength coverage of the SDSS spectra we were able to
search for the redshifted  bump in 8 quasars of our list.

In 7 of these   quasars  
we do not find evidence of the bump
from the comparison of the quasar spectrum with   
the MW-type extinction curve (CCM with $R_V=3.1$)
 plotted in a scale $\nobreak{10^{-\xi[\lambda/(1+z_\mathrm{abs})]} }$
 which, apart from a constant factor, represents the transmission of  the DLA system  
 (see Eq. \ref{newA6} in Appendix A).
One example of negative detection
is shown in Fig. \ref{bump}a, where one can see that
the continuum of J0016-0012 changes smoothly,
as expected by a power law, in
the region where the bump is   expected to produce a
broad, deep absorption.
We cannot exclude that, in exceptional cases, a particular combination of quasar
emission lines may conspire to hide a dust bump, if present.
However, it is unlikely that this happens systematically in the 7 cases
considered here, characterized by different combinations of emission
and absorption redshifts.  
In particular, 
we do not find the bump in J1159+0112 and J0016-0012,
the two cases for which  the test of Fig. \ref{AVtest}
indicates an SMC type extinction.
In these two cases, therefore, we have 
two tests, one based on photometry and the other on spectroscopy, that consistently 
suggest an SMC-type extinction curve.

Only in one case, J0121+0027, 
we found tentative evidence of the bump.
This case is shown in Fig. \ref{bump}b, where one can see that the continuum
of J0121+0027 runs smoothly at $\lambda > 600 \, \mathrm{nm}$, but shows
an absorption at $400 \, \mathrm{nm} \lsim \lambda \lsim 600 \, \mathrm{nm}$.
 The position, width and intensity of this absorption feature
are broadly consistent with those predicted for the   bump,
given the uncertainties induced by the quasar emission lines.
For this absorber we 
consider the possibility that the extinction curve  
may be of Milky-Way type, without ruling out completely  a SMC-type curve (see Table 3). 
We refer to Wang et al. (2004) for a detailed spectroscopic analysis
of this case.

For the absorbers of our list without spectral coverage of the bump
we adopted an SMC curve. 
This choice is justified by the results of Wild \& Hewett (2005) and York et al. (2006)
which have found no evidence of the bump in their average absorber spectra. 
In any case,
the results for an SMC and MW curve are  often very similar.
For the DLA system in front of   J1323$-$0021,
the only reddened quasar of our list without adequate spectral coverage\footnote
{
If present, the  bump of the absorber at 
$z_\mathrm{abs}=0.716$ in J1323$-$0021 should fall at the violet edge of
the SDSS spectrum.  At shorter wavelengths,  the Ly\,$\alpha$ emission of the quasar
and   the Ly\,$\alpha$ forest will make difficult the search for     
  the bump even in spectra with better coverage.
}
of the bump,
we expect a difference of only 0.08 dex in  $A_V$.

The resulting values of extinction are listed in Table 3. 
The  large upper error bar that we assign to J0121+0027 reflects
the uncertainty in the choice of $R_V$,
since a value as high as $R_V \sim 5.5$ has been proposed by Wang et (2004)
for the absorber at $z_\mathrm{abs}=1.388$ (see footnote $h$). 
If  $R_V$ is so high,
the extinction curve becomes flat and
the conversion from reddening to $A_V$ unreliable, as we
explained above. 
In any case the value of extinction inferred by Wang et al. for this system,
$\log A_V \simeq +0.10$ dex, lies
well within our upper error bar.

\begin{center} 
\begin{table*} 
\scriptsize{
\caption{Extinction and metal column densities of the DLA systems of Table 1.}
\begin{tabular}{lcccccccc}
\hline \hline 
 & & & & & & & &\\
~~ SDSS  &    $z_\mathrm{abs}$  &  $\langle \log A_V \rangle^a$ &  $(\log A_V)^b_\mathrm{pred}$ & $\log N(\ion{Zn}{ii})^c$ & ~ [Zn/Fe]$^c$
& $(\log \widehat{N}_\mathrm{Fe})^d$ & $(\log \widehat{N}_\mathrm{Fe})^e$ & $(\log \widehat{N}_\mathrm{Fe})^f$ \\
&  & & & & & & &\\
\hline
&  & & & & & & &\\
J0013+0004  & 2.025   & $< -1.02$ & $-1.89,-1.67$& $12.25 \pm 0.05$ & $0.05\pm 0.07$      & $< 15.00^g$	   & $< 15.00^g$ &   $<14.80^g$   \\
J0016-0012  & 1.973   & $-0.80_{-0.21}^{+0.11}$ & $-1.22,-1.00$& $12.82 \pm 0.04$ & $0.84 \pm 0.05$     & $15.61 \pm 0.04$ & $15.62 \pm 0.04$ &  $15.50 \pm 0.04$ \\
J0121+0027$^h$  & 1.388   & $-0.16_{-0.15}^{+0.76}$ & --- &$>13.32$ &  $> 1.48$      & $> 16.17$	   & $> 16.44$  &  $>16.19$   \\
J0938+4128  & 1.373   & $< -0.69$ & $-1.85,-1.63$&$12.25 \pm 0.05$ & $0.29 \pm 0.11$     & $14.80 \pm 0.14$ & $14.80 \pm 0.14$ & $14.56 \pm 0.14$ \\ 
J0948+4323 & 1.233 & $< -0.51$ & $-1.03,-0.80$&$13.15 \pm 0.01$ & $0.45 \pm 0.02$   & $15.82 \pm 0.02$ & $15.82 \pm 0.02$ & $15.65 \pm 0.02$ \\ 
J1010+0003  & 1.265 &$< -0.88$ &$-0.98,-0.76$ &$13.15 \pm 0.06$ & $0.75 \pm 0.08$ &$15.93 \pm 0.06$ & $15.93 \pm 0.06$ & $15.80 \pm 0.06$ \\
J1107+0048  & 0.741   & $< -0.58$ & $-1.01,-0.79$&$13.03 \pm 0.05$ & $0.37 \pm 0.08$     & $15.65 \pm 0.08$ & $15.65 \pm 0.08$ & $15.46 \pm 0.11$ \\
J1159+0112  & 1.943   & $-0.85_{-0.22}^{+0.10}$ & $-1.25,-1.02$&$13.09 \pm 0.08$ & $0.45 \pm 0.11$     & $15.76 \pm 0.10$ & $15.76 \pm 0.10$ & $ 15.59 \pm 0.12$ \\
J1232$-$0224  & 0.395   & $< -0.50$ &$-1.10,-0.87$ &~$12.93 \pm 0.12$ & ~$0.83 \pm 0.16^k$ & $15.72 \pm 0.12$ & $15.72 \pm 0.12$ & $15.60 \pm 0.13$ \\
J1323$-$0021 & 0.716   & $ -0.36_{-0.12}^{+0.07}$ & $-0.56,-0.33$&$13.43 \pm 0.05$ & $1.14 \pm 0.06$    & $16.26 \pm 0.05$ & $16.28 \pm 0.05$ & $16.16 \pm 0.05$ \\ 
J1501+0019 & 1.483 &$< -0.79$ &$-0.92,-0.70$ &$13.10 \pm 0.05$ & ~$0.40 \pm 0.09^k$ &$15.74 \pm 0.08$ & $15.74 \pm 0.08$& $15.56 \pm 0.11$ \\
J2234+0000  & 2.066   & $< -0.61$ &$-1.60,-1.38$ &$12.46 \pm 0.02$ & $0.57 \pm 0.06$     & $15.19 \pm 0.03$ & $15.19 \pm 0.03$ & $15.05 \pm 0.04$\\
J2340$-$0053  & 1.360   & $-0.67_{-0.32}^{+0.11}$ & --- &$12.62 \pm 0.05$ & $0.54 \pm 0.13$   & $15.33 \pm 0.07$ & $15.33 \pm 0.07$ & $15.18 \pm 0.10$ \\
&  & & & & & & &\\
\hline
\end{tabular}
\scriptsize{  
\\
$^a$ Measured from the observed reddening using Eq. (\ref{NumSol}) for the conversion to rest frame extinction.
\\
$^b$ Extinction predicted from Eq. (11) of Vladilo \& P\'eroux (2005) for $G=0.6$   and $G=1$, respectively.
\\
$^c$ See references in Table 1.
\\
$^d$  Calculated assuming Zn completely undepleted and   [Zn/Fe]$_a=0$. 
\\
$^e$  Calculated using a scaling law of interstellar depletions 
(Vladilo 2002b; $\epsilon_\mathrm{Zn}=1$, $\eta_\mathrm{Zn}=42.5$) and [Zn/Fe]$_a=0$.  
\\
$^f$ As in the previous column, but with [Zn/Fe]$_a=+0.1$ dex. 
\\
$^g$ Dust-free system; upper limit computed introducing $\log N(\ion{Zn}{ii}) < 12.40$ 
and $\log N(\ion{Fe}{ii}) > 14.91$  in Eqs. (\ref{NFedust1}) and (\ref{NFedust2}).
\\
$^h$ 
Central and upper values of $\langle \log A_V \rangle$ calculated using a CCM model with $R_V=3.1$ and $R_V=5.5$, respectively.
 Lower value calculated using the SMC extinction curve (see Section 3.1).
\\
$^k$ $N$(Ni) or $N$(Cr) used as a proxy of $N$(Fe), which is not measured;
 Ni and Cr are good tracers of Fe in stars with   [Fe/H] $\gsim -2$ dex (e.g. Ryan et al. 1996);
Ni, Cr and Fe have very similar interstellar depletions  (e.g. Vladilo 2002a);
we adopt [Ni/Fe]=0 and [Cr/Fe]=0 in the conversion.
}
}
\end{table*}
\end{center}
\normalsize

\section{Extinction versus metal column density}

On the basis of the  
of the selection process of our sample (Section 2)
we assume that  the DLA systems of   Table 1
are the major source of reddening of their   quasars
and  we investigate the relation 
between the quasar extinction  
and the DLA metal column density. 
  The rest-frame extinction scales  
with the  dust-phase column density of iron,  
  $\widehat{N}_\mathrm{Fe}$ [cm$^{-2}$],  
 according to the relation
\begin{equation}
A_V  = 
\,\, \langle s^\mathrm{Fe}_V \rangle ~ \widehat{N}_\mathrm{Fe}  ~,
\label{DeltaNorMetCol}
\end{equation}
derived in Appendix B. The term
$\langle s^\mathrm{Fe}_V \rangle $ is the mean optical cross section 
of the dust grains in the $V$ band per atom of iron in the dust. 
The   extinction  will change  between different  lines of sight
   tracking the variations of the   dust-phase column density of iron
and of the mean cross section $\langle s^\mathrm{Fe}_V \rangle $.
Our goal is to  probe the behaviour of $\langle s^\mathrm{Fe}_V \rangle $   in different \ion{H}{i} regions
from a simultaneous measurement of  $A_V$
and $\widehat{N}_\mathrm{Fe}$.
The measurement of $\widehat{N}_\mathrm{Fe}$
in extragalactic clouds is discussed in the next section.

\subsection{The dust-phase column density of iron}
 
The  dust-phase column density of iron  
cannot be measured directly from optical/UV spectroscopic data. 
With this type of observations,
however, we can infer  $\widehat{N}_\mathrm{Fe}$
from the  measured gas-phase column densities of zinc and iron.  
We use zinc as gas-phase tracer of iron, since zinc is a volatile element
with little affinity to dust (Savage \& Sembach 1996)
and, at the same time, the zinc/iron abundance
ratio is approximately constant, in solar proportion, 
in a large fraction of Galactic stars over a very wide range of metallicities
(Mishenina et al. 2002, Gratton et al. 2003, Nissen et al. 2004). 

If zinc were completely undepleted and the intrinsic Zn/Fe ratio in DLA systems perfectly solar,
the Zn/Fe ratio observed in the gas would give an exact measure of the iron depletion
$\delta_\mathrm{Fe} \equiv
\log (N_\mathrm{Fe}/N_\mathrm{Zn}) - 
\log \mathrm{(Fe/Zn)}_{\sun}$. 
In this case the  dust-phase column density of iron would be      
\begin{equation}
\widehat{N}_\mathrm{Fe} 
=  f_\mathrm{Fe} 
\,   N_\mathrm{Zn}  \,
 \left( { {\mathrm{Fe} \over \mathrm{Zn} } } \right)_{\! \sun}  
 ~,
\label{NFedust2}
\end{equation} 
where  $f_\mathrm{Fe} =
  1-10^{\delta_\mathrm{Fe}} $ is the fraction of iron  in the dust. 
With this expression $\widehat{N}_\mathrm{Fe} $
can be directly estimated from the measured column densities
$N_\mathrm{Zn}$ and $N_\mathrm{Fe}$. 
 
A closer inspection of interstellar data, however,
indicates that zinc also can be depleted, with an extreme value
 as high as $-0.6$ dex 
in dense clouds such as that in front of $\zeta$\,Oph (Savage \& Sembach 1996).
Therefore, we must take into account the possibility
that   zinc  may also be depleted (in small amounts) in DLA systems. 

In addition, a closer analysis of stellar data shows that the Zn/Fe ratio 
 tends to increase with decreasing metallicity.
The effect is strong when [Fe/H] $< -2$ dex,
attaining a value as high as [Zn/Fe] $\simeq +0.5$ dex at [Fe/H] $\sim -4$ dex
(Cayrel et al. 2004). Luckily, this interval of metallicities is well below
the typical metallicities of DLA systems.
However, a mild excess has also been found in a fraction of Galactic stars 
which are less metal deficient (Prochaska et al. 2000, Chen et al. 2004).
For stars of the thin disk the typical excess is 
[Zn/Fe] $\simeq +0.1$ dex at [Fe/H] = $-0.6$ dex
and vanishes at higher metallicities  (Chen et al. 2004). 

By allowing a fraction $f_\mathrm{Zn}$ of zinc  
to be  in the dust and the intrinsic Zn/Fe ratio
of the absorbers, $\left( {\mathrm{Zn} \over \mathrm{Fe} }\right)_{a}$, 
to deviate from the solar value, we obtain the more general expression  
\begin{equation}
\widehat{N}_\mathrm{Fe} 
=   f_\mathrm{Fe}  \,  { N_\mathrm{Zn} \over   (1-f_\mathrm{Zn}) } \,
\left( {\mathrm{Zn} \over \mathrm{Fe} }\right)^{-1}_{a}
 ~.
\label{NFedust1}
\end{equation} 
%
With an educated guess of $\mathrm{(Zn/Fe)_{a}}$
it is possible to estimate $\widehat{N}_\mathrm{Fe}$
from the  column densities
$N_\mathrm{Zn}$ and $N_\mathrm{Fe}$
also in this general case,
assuming that $f_\mathrm{Zn}$  scales with $ f_\mathrm{Fe}$ according to
a scaling law of interstellar depletions calibrated in the ISM. 
Details on this method can be found in Vladilo (2002a, 2002b).  
In Table 3 we 
give the   dust-phase column densities of iron
calculated
for three different cases:
(1) zinc completely undepleted and $\mathrm{[Zn/Fe]_{a}}=0$
(column 7);
(2) zinc depletion scales with iron depletion and $\mathrm{[Zn/Fe]_{a}}=0$
(column 8);
(3)   zinc depletion scales with iron depletion and $\mathrm{[Zn/Fe]_{a}}=+0.1$
(column 9). 
   \begin{figure*}
   \centering
 \includegraphics[width=12 cm,angle=0]{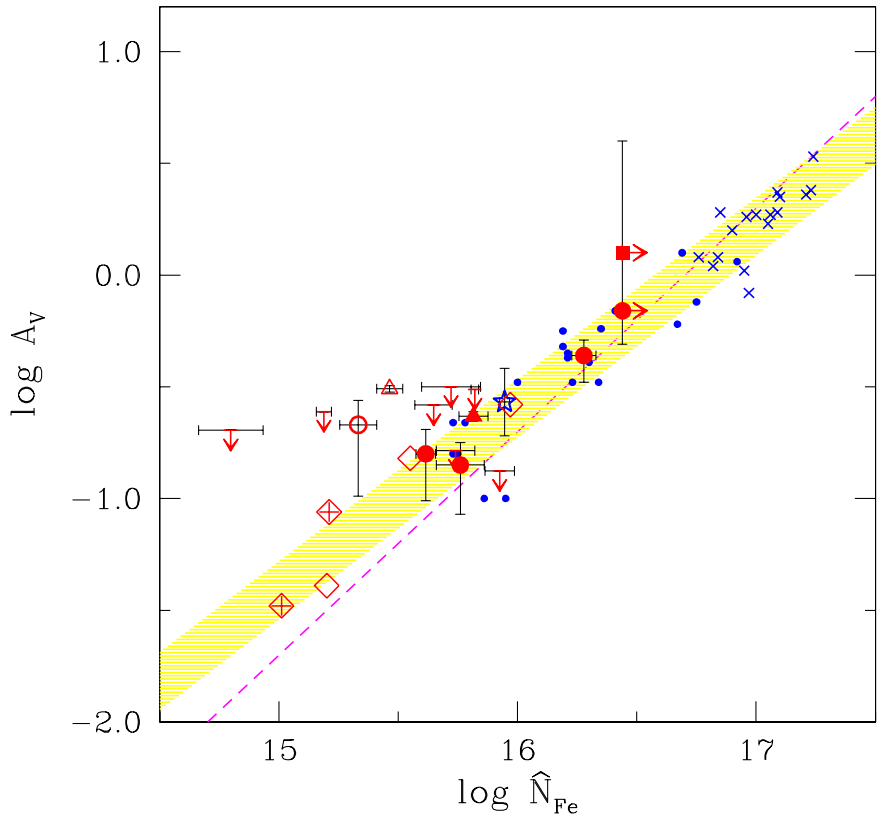}  
  \caption{    Extinction, $A_V$,    versus  dust-phase column density of iron,
  $\widehat{N}_\mathrm{Fe}$,  in QSO absorbers and interstellar clouds. 
  The extinction is calculated in the rest frame of the absorber.  
Circles  and arrows pointing down: $A_V$ measurements
and upper limits from this work; empty circle: J2340$-$0053.  
Triangles: QSO/DLAs from Ellison et al. (2005); empty triangle: B0438$-$436.
Square: extinction for J0121+0027 from Wang et al. (2004). 
Crossed diamonds: sub-samples of \ion{Mg}{ii} absorbers from York et al. (2006). 
Open diamonds: sub-samples of \ion{Ca}{ii} absorbers from Wild et al. (2006).
  Open star: SMC line of sight towards Sk 155 (Welty et al. 2001, Fitzpatrick 1985). 
  Dots and crosses:  Milky Way interstellar lines of sight
  from Jenkins et al. (1986) and Snow et al. (2002), respectively. 
  Strip: $\pm 1 \, \sigma$ interval around the linear regression
  of the MW data. 
  Dashed line: linear regression of the MW data with
  fixed unit slope.  See Sect. 4.2. }
              \label{AV-NdFe}%
    \end{figure*}

The comparison between  columns 7 and 8 indicates that the results 
are almost unaffected by the possible depletion of zinc.  
Only when the depletion level is particularly high,
i.e. when [Zn/Fe] $\gsim$ 1 dex, the correct relation (\ref{NFedust1})
yields slightly higher values of $\widehat{N}_\mathrm{Fe}$
than the approximate relation (\ref{NFedust2}). 

The comparison between  columns 8 and 9 indicates that
a mild excess of the intrinsic abundance ratio
$\mathrm{[Zn/Fe]_{a}}$ yields a similar mild
decrease of $\widehat{N}_\mathrm{Fe}$ with respect to the case
$\mathrm{[Zn/Fe]_{a}}=0$. 
The higher the observed [Zn/Fe] ratio, the lower the
difference. For the quasars with reddening detection
the  observed [Zn/Fe] ratio of the DLA 
 happens to be relatively high
(column 4 in Table 3).
Therefore the results of the present work are weakly affected
by an excess of  $\mathrm{[Zn/Fe]_{a}}$ of the type observed
in stars of  moderately low metallicity.

In fact, for two of the quasars with reddening detection
(J0121+0027 and J1323$-$0021) the metallicity of
the intervening DLA system is relatively
high (Wang et al. 2004, P\'eroux et al. 2006)
and we do not expect an excess of $\mathrm{[Zn/Fe]_{a}}$
 given the trend of  
[Zn/Fe]  
observed in Galactic stars (Chen et al. 2004).
The absorber with lowest metallicity among the
5 quasars with reddening detection is that
 at $z_\mathrm{abs}=1.944$ towards J1159+0012
([Zn/H] $\simeq -1.4$ dex). 
This is the absorber in which   $\mathrm{[Zn/Fe]_{a}}$
 could be, in principle, most enhanced. As we show below,
the results of this paper are unaffected
even in this case.

\subsection{The empirical relation between $A_V$
and $\widehat{N}_\mathrm{Fe}$}

In  Fig. \ref{AV-NdFe} we plot the rest frame extinction,  $A_V$,
versus the dust-phase column density of iron, $\widehat{N}_\mathrm{Fe}$,
 for different types of absorbers of the local and high redshift Universe.

The data obtained from our reddening detections  
at the $\nobreak{2 \, \sigma}$ level
are indicated with circles. 
The filled circles are the quasars in which the DLA system of Table 1
is the major source of reddening.  
The empty circle is J2340-0053, for which  
an additional contribution to the reddening may be present.
 For J0121+0027 we also plot the extinction derived by Wang et al. (2004)
from the comparison with composite SDSS quasar spectra
(filled square).

For the cases without reddening detection  
we plot  a 2\,$\sigma$ upper limit. 
The values of $\widehat{N}_\mathrm{Fe}$ 
are estimated from Eq. (\ref{NFedust1}), case [Zn/Fe]$_a$=0.
The implications of these choices are discussed below. 

The triangles in the figure  represent  two quasars
 with     photometric 
 reddenings and DLA metal column densities 
obtained in the framework of the CORALS survey (Ellison et al. 2005). 
These quasars are not included in SDSS catalog. 
To estimate $A_V$  we adopted the normalized colour $(B-K)_n$
in Table 3 of Ellison et al. and an SMC extinction curve.  
Only two quasars of this survey (B0438$-$436 and B0458$-$020)
have a significant $(B-K)$ reddening 
and, at the same time,
Zn and Fe column densities (Akerman et al. 2005).
The quasar B0438$-$436 is represented with an empty triangle,
rather than a filled one, 
because its reddening seems to be contaminated by dust
in the quasar environment  (Ellison et al. 2005).

The crossed diamonds in the figure are  taken 
from a study 
of the average reddening properties
of 809 \ion{Mg}{ii} systems in SDSS quasar spectra (York et al. 2006). 
The redshift interval covered by this study is similar to that of the DLA
systems of Table 1. However,   \ion{Mg}{ii} absorbers may have, in general,
\ion{H}{i} column densities well below the DLA regime.
For  comparison with our data we have considered the
sub-samples  of York et al.
with largest equivalent widths of \ion{Mg}{ii}  because
in these sub-samples the fraction of DLA systems is expected
to be the highest.  
The results shown in the figure represent the 
mean values of $A_V$ and $\widehat{N}_\mathrm{Fe}$ for
the sub-samples S8 
(251 systems with equivalent widths $>$ 2 \AA)
and S26
(97 systems with equivalent widths $>$ 2.5 \AA\ and   $\Delta(g-i) < 0.2$ magnitudes).
Sub-samples with saturated \ion{Zn}{ii} and \ion{Fe}{ii} mean profiles were not considered.
The mean extinction was estimated from the mean $E(B\! - \!V)$ of 
these sub-samples
using the SMC value $R_V=2.74$ (Gordon et al. 2003). 
The mean $\widehat{N}_\mathrm{Fe}$  was derived by inserting
the mean $N(\ion{Zn}{ii})$ and [Zn/Fe] 
of the sub-samples in relation (\ref{NFedust1}). 

The open diamonds  are  taken from the study  
of  \ion{Ca}{ii} systems at $z \sim 1$ in SDSS quasar spectra by Wild et al. (2006). 
In order of increasing extinction, they represent the
mean values of $A_V$ and $\widehat{N}_\mathrm{Fe}$ for
the sub-samples  labelled Low-$W_{\lambda3935}$, All, and
High-$W_{\lambda3935}$,
respectively.
We estimated these  mean values  
  from the corresponding mean values
of $E(B\! - \!V)$, $N(\ion{Zn}{ii})$ and [Zn/Fe] given by Wild et al. (2006)
for the     27 absorbers analysed for element column densities. 
Also in this case we assume an SMC type of dust.

The quasar absorption data in  Fig.   \ref{AV-NdFe}
are consistent with an increase of $A_V$  with 
increasing $\widehat{N}_\mathrm{Fe}$.
A linear trend between these two quantities is
predicted by relation (\ref{DeltaNorMetCol})
if the mean extinction per atom of iron
is approximately constant in \ion{H}{i} interstellar regions.
However, the number of data points 
for the individual quasar absorbers (circles and triangles) 
is too little for deriving a statistical correlation.
In order to cast light on the behaviour of high redshift clouds
we added in the same figure  
results obtained for local interstellar clouds.

The  dots and crosses 
in the figure represent Milky-Way
lines of sight with available measurements
of $A_V$, $N(\mathrm{H_{tot}})=N(\ion{H}{i})+2 N(\mathrm{H}_2)$ and $N(\ion{Fe}{ii})$,
taken from Jenkins et al. (1986) and Snow et al. (2002), respectively.
For these data   we derived
$\widehat{N}_\mathrm{Fe}$
assuming that the local interstellar abundance of iron is solar.  
The error budget in the estimate of $\widehat{N}_\mathrm{Fe}$ is dominated
by the error of  $N(\mathrm{H_{tot}})$, which is typically of $\sim 0.1$ dex.
The extinctions were taken from the catalog of Neckel \& Klare (1980),
under the entry $A_V$(MK).
Uncertainties of these $A_V$ values are in the order of 0.1 mag.
Values with $A_V < 0.1$ mag  were not considered.  
Lines of sight with $N(\ion{H}{i})$ below the definition threshold of DLA systems 
and with saturated iron lines were rejected. 

The Milky-Way data are highly correlated, with linear correlation coefficient $r=0.95$
and slope $m=0.81 \pm 0.27$.  
The strip in the figure represents the dispersion of the data
around the linear regression. 
The dashed line is the linear regression with fixed $m=1$, a slope
consistent with the free regression analysis. 

 The star symbol in the figure  represents   
the SMC star Sk 155,
the only Magellanic sightline  for which we could determine  
$A_V$ and $\widehat{N}_\mathrm{Fe}$.
The extinction was estimated using the SMC value
of $E(B\! - \!V)$ from Fitzpatrick (1985) and
the mean SMC value $R_V=2.74$ (Gordon et al. 2003).  
We estimated
$\widehat{N}_\mathrm{Fe}$ from relation 
(\ref{NFedust1}) using the Zn and Fe column densities 
of Welty et al. (2001)
integrated over the SMC radial velocities. 
In spite of the differences in metallicity level and type of
extinction curve relative to the Milky Way, the data point of  Sk 155
is in excellent  agreement with the Milky-Way data.

Most of the DLA measurements (filled circles and triangles) follow 
the trend of   Milky-Way interstellar data.  
Uncertainties in
the choice of the extinction curve 
are unlikely to alter this result. 
Even in the case of J0121+0027,
for which these uncertainties are large,  
the DLA data point seems to be consistent with the interstellar trend
(the value of $\widehat{N}_\mathrm{Fe}$ is a lower limit
owing to saturation of the \ion{Zn}{ii} lines). 

Also the  uncertainty in the adopted value of  $\nobreak{\mathrm{[Zn/Fe]}_a }$
in Eq. (\ref{NFedust1}) is unlikely to affect the agreement 
between DLA systems and interstellar clouds.  
As we said above, 
J1159+0112 is the 
most likely candidate for  
a mild enhancement of $\nobreak{\mathrm{[Zn/Fe]}_a }$
among our quasars with reddening detection. 
By adopting
an  enhanced [Zn/Fe]$_\mathrm{a}$ in this case 
(last column of Table 3), we
still find a good agreement with the Milky-Way data.

In spite of this general agreement,
an excess of extinction relative to the  Milky-Way trend
is found in J2340-0053 (empty circle)
and B0438-436 (empty triangle).  
Quite interestingly, these are exactly the cases for which
we suspect the presence of extinction sources in addition
to the DLA system. 
Therefore, the excess may be due to the
presence of additional absorbers rather than to
a deviation of individual DLAs from the interstellar trend.

Also the $A_V$ upper limits  that we derive  
are  generally consistent with the interstellar data. 
This conclusion would be strengthened by adopting
an enhanced [Zn/Fe]$_a$ in Eq. (\ref{NFedust1}),
instead of $\nobreak{\mathrm{[Zn/Fe]}_a = 0}$,
since in this case the limits would shift slightly to the left (see Table 3).

In summary, the data collected in the figure suggest that the trend between
$A_V$ and $\widehat{N}_\mathrm{Fe}$  is remarkably similar
in interstellar clouds of the Milky Way and the SMC and in 
DLA systems with different
metallicities (from $\simeq -1.4$ dex up to $\simeq +0.5$ dex relative
to solar) and redshifts ($z \simeq 0.7$ up to $z \simeq 2$). 
 Also \ion{Ca}{ii} absorbers and the  \ion{Mg}{ii}  absorbers with highest values 
of equivalent width follow the same trend. 

In the interpretation of the results of Fig.   \ref{AV-NdFe} one must take into account
that some parts of the plot are not   accessible to the observations.
For instance, if $A_V$ is too large, the background quasar  drops
out of the SDSS sample. If it is too low, the reddening cannot be detected.
 The detection limit, in addition,  will vary as a function of the redshift
and of the scatter  of the intrinsic colors of the quasars.
These effects  may alter the distribution of detected DLA systems
 in Fig.   \ref{AV-NdFe}
   and,  in principle,   could
induce an artificial trend among the data points.

The agreement between the different sets
of high redshift data 
in Fig.   \ref{AV-NdFe}
is impressive,
considering the different methods employed for measuring
the reddening.
Also the agreement between high redshift and local data     
  is remarkable, considering
the different methods used to derive $A_V$ and $\widehat{N}_\mathrm{Fe}$ 
(for instance,
the educated guess of [Zn/Fe]$_a$ was applied 
to derive $\widehat{N}_\mathrm{Fe}$ at high redshift, 
but not in the Milky Way).  
The general agreement between DLA  and interstellar data speaks against
the existence of an artificial  trend induced by selection effects
of DLA detections. 
The implications 
  of this general agreement  are discussed below.

\section{Discussion}
 
\subsection{Implications for the properties of dust grains}

The existence of a common trend between $A_V$ and $\widehat{N}_\mathrm{Fe}$
in   \ion{H}{i} regions of galaxies with different metallicities
and redshifts  indicates that the dust grain parameter
\begin{equation}
\langle s^\mathrm{Fe}_V \rangle
=   1.007 \! \times \! 10^{-22} ~
{   \sum_j w_j \,  \langle \, Q_{\lambda_V} ~ \sigma_g \, \rangle_j
\over
\sum_j w_j \,  \langle  \,V_g \, \varrho  ~ X_\mathrm{Fe}  \, \rangle_j
} 
\label{sFelPAP}
\end{equation} %
representing the mean extinction  in the $V$ band
per atom of iron in the dust
(see Appendix B), is approximately constant
in a large variety of neutral regions.  
The scatter of the Milky Way data is   of only 0.16 dex at 1\,$\sigma$ level. 
The few measurements in DLA systems do not show, so far, evidence
for a larger dispersion.

In principle,  the $V$ band could be less sensitive    than   other spectral bands
to variations of the extinction properties of the grains.
Notwithstanding,
the small scatter of $\langle s^\mathrm{Fe}_V \rangle$
is rather remarkable given the fact
that the value of this parameter  is determined
by at least 4 different properties of the grains:
their size, $a$, internal density, $\varrho$, abundance by mass of iron
inside the grain, $X_\mathrm{Fe}$, and extinction efficiency factor,
$Q_\lambda$ , which is related to the geometrical and optical properties
of the grains (Spitzer 1978).
We know that these parameters
    vary among interstellar clouds. 
In particular, the   variations of the extinction curves
 are   attributed, in large part,
to variations of the grain size distribution (Draine 2003). 
Therefore, in order to produce an approximately constant $\langle s^\mathrm{Fe}_V \rangle$
there must be a physical mechanism able to compensate the variations
of the individual parameters that appear in the above expression.
We tentatively propose the following mechanism
 based on heuristic considerations.

We consider two types of interstellar environments:
(1) regions where grain destruction mechanisms are efficient
and
(2) clouds where the grains are protected from destruction. 
In the second case we expect 
that the grain size $a$ can be larger, on the average, and the  volatile elements more
easily incorporated into the grains
than in the first case. 
The relative abundance of iron
inside the grains, $X_\mathrm{Fe}$,
must be lower in the second case, when also
volatile elements are incorporated in the grains, than in the first case,
when refractory elements, such as iron, are dominant. 
Since the most abundant volatile elements (e.g. carbon)
are lighter than the most abundant refractory elements (e.g. iron)
we expect, in addition, that in the second case  the
mean density $\varrho$ is lower than in the first case.

Combining all together, we expect  
the following variations  of the grain parameters to occur  
passing from a grain-destructive to a  non-destructive environment: 
an increase of $a$ and, at the same time, a decrease of $\varrho$ and $X_\mathrm{Fe}$.
From relation (\ref{sFelPAP}) we expect that
$\langle s^\mathrm{Fe}_V \rangle \approx (a \, \varrho \, X_\mathrm{Fe})^{-1}$
since $\sigma_g \approx a^2$ and $V_g \approx a^3$. 
Given this dependence of $\langle s^\mathrm{Fe}_V \rangle$,
  the changes of $a$ could compensate
the changes of $\varrho$ and $X_\mathrm{Fe}$.

A much larger database of   interstellar data at low and high redshift are required to
understand the viability
of the simple mechanism of compensation proposed here.
Contraints on this mechanism could be obtained   by 
studying   the slope of the    $A_V$ and $\widehat{N}_\mathrm{Fe}$ relation
and analysing possible variations of $\langle s^\mathrm{Fe}_V \rangle$ versus $R_V$.
Given the paucity of quasar absorption extinction measurements 
these studies are still premature at high redshift, but could be the subject of 
future investigations.

\subsection{Implications for the dust obscuration bias}

The extinction of the quasars due to the intervening DLA systems
is expected to produce a selection effect in optical, magnitude-limited surveys:
the absorbers with highest values of extinction would be systematically missed from the
surveys  as a consequence of quasar  obscuration (Fall \& Pei 1989). 

Evidence for this effect has been searched by comparing the statistics
of optical surveys and radio-selected surveys, the latter being unaffected by the
extinction. 
A marginal signature of the bias has been found in this way (Ellison et al. 2001, Akerman et al. 2005), 
but the statistics are still based on small numbers and the estimate of the effect
uncertain due to the difficulty of observing 
all the quasars of the radio-selected sample, independently of their magnitude,
in the optical follow-up.

In a previous work we have proposed a new approach for quantifying the effect of quasar
obscuration on the statistics of DLA systems. 
Using a relation between the extinction and the metal column densities of the absorbers
we  invert the frequency distributions of \ion{H}{i} column densities and metallicities
measured from magnitude-limited surveys and derive the unbiased frequency distributions
(Vladilo \& P\'eroux 2005). 
The existence of a general relation between extinction and metal column density
is fundamental in this type of approach. 
The results of the present work  
  lend quantitative support to the 
relation  adopted in our previous work\footnote{
See Eq. (11) in Vladilo \& P\'eroux (2005), where 
$\nobreak{Z \equiv N_\mathrm{Zn} / N_\mathrm{H} }$
and  $G=1$ ($0.6$) for MW (SMC) type of dust;
$\xi(\lambda_V) \equiv 1$ in the rest frame of the absorber;
$\nobreak{
f_\mathrm{Fe}(Z)=
{ 1 \over \pi}  \,
\{
 \arctan
\left({\mathrm{[Zn/H]} + 1.25 \over 
0.4} \right)+{\pi \over 2} 
\}
\, .
}
$ 
Apart from a normalization factor,
the parameter $G$ is equivalent to $\langle s^\mathrm{Fe}_V \rangle$
for the special case of a single family of identical, spherical grains.
}, 
$\nobreak{A_V \simeq 1.85 \times 10^{-14} \, G \, f_\mathrm{Fe}(Z) \, N_\mathrm{H} \, Z}$, 
which was mostly based on interstellar data. 
In Table 3, Col. 4, we list the values of extinction predicted from this previous relation
for the DLA systems of the present work.
One can see that the predictions are  in most cases consistent
with the measurements or upper limits.
The agreement with the measurements is better for $G=1$ than
for $G=0.6$ (in this latter case the extinction is mildly underestimated).  
These results indicate that the quantitative predictions of the obscuration bias
presented by Vladilo \& P\'eroux (2005)
are realistic and, in any case, not overestimated.

In order to confirm the general validity of the relation between extinction
and metal column densities in DLA systems, it would be important to obtain
more measurements at higher redshifts since the present sample is limited
at $z_\mathrm{abs} \simeq 2$, while
 a large fraction of DLA systems is observed to be at higher redshift.
In particular, it will be crucial to calibrate the relation at high values
of extinction ($A_V > 1$ mag), when the obscuration bias is most critical. 
The slope $m \simeq 0.8$ of the interstellar data is consistent with a
mild decrease of the extinction per atom of iron in the dust
with increasing metal column density (strip in Fig. \ref{AV-NdFe}). 
For an exact estimate of the obscuration bias it is critical to verify 
the existence of this effect in DLA systems.

   \begin{figure*}
   \centering 
    \includegraphics[width=8.5 cm,angle=0]{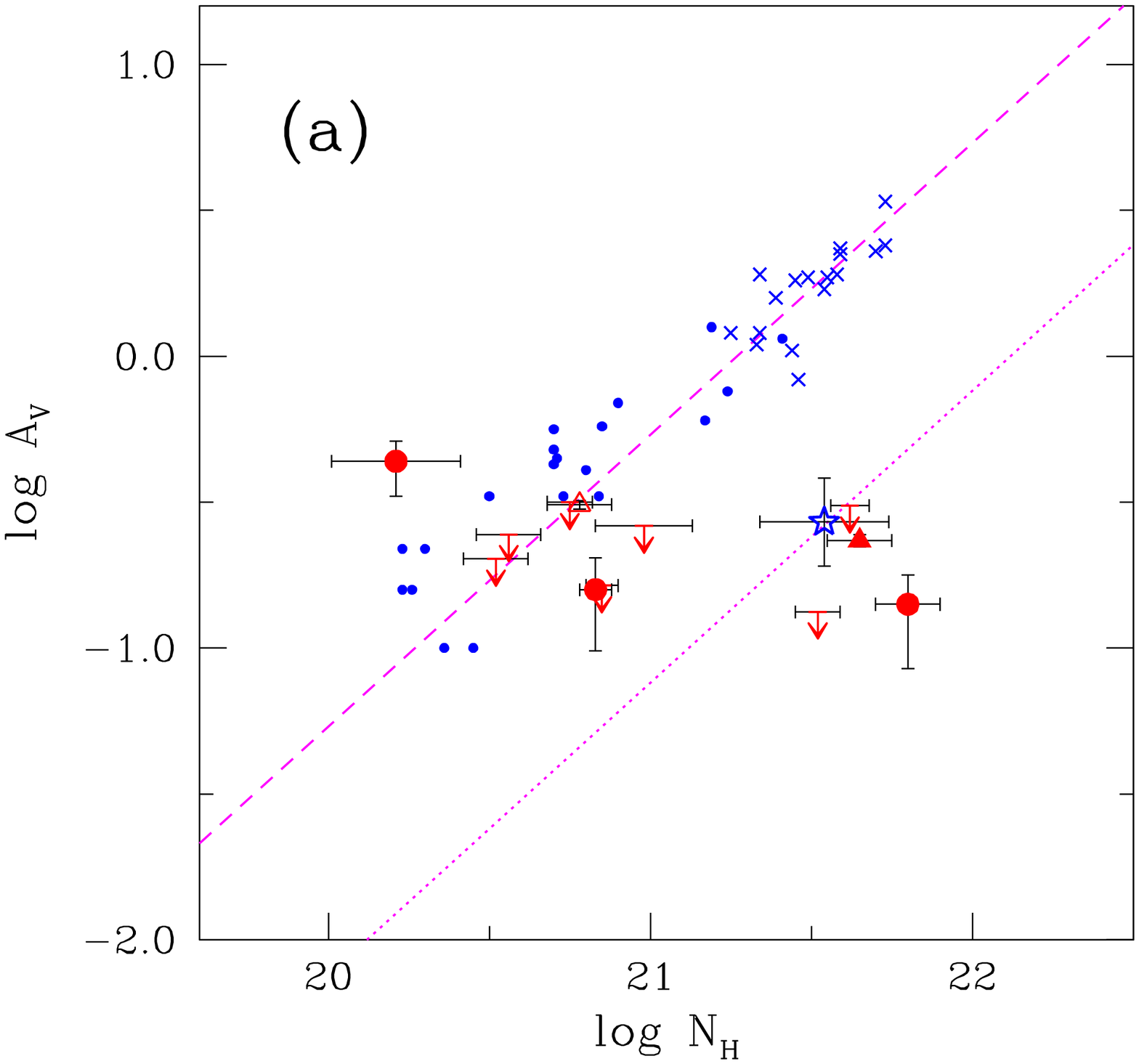} 
    \hskip 0.5cm
     \includegraphics[width=8.5 cm,angle=0]{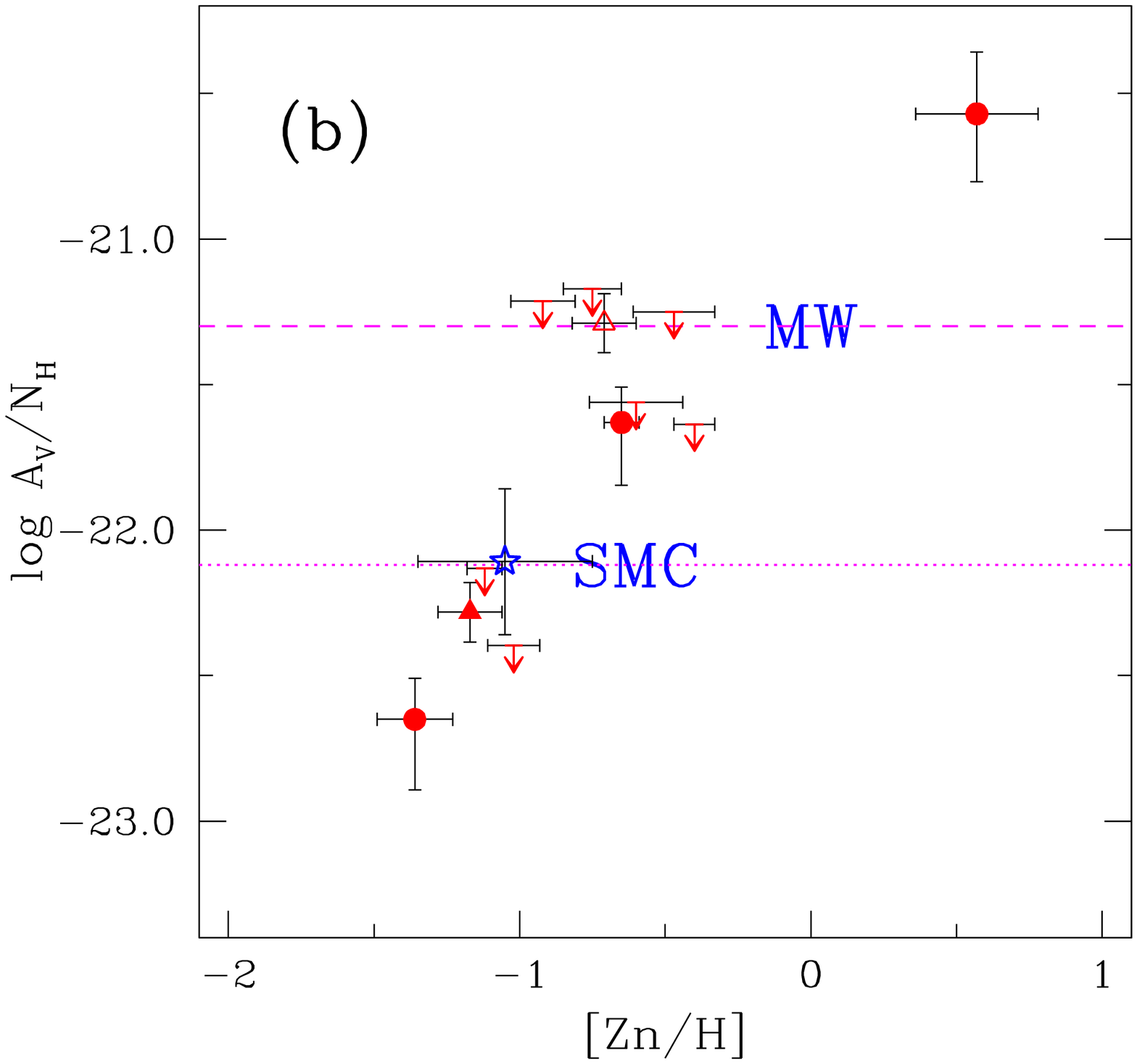} 
  \caption{    
   {\bf (a)} Extinction, $A_V$, versus total hydrogen column density,  $N_\mathrm{H}$,
   of DLA systems and interstellar clouds. Same symbols as in Fig. \ref{AV-NdFe}.
    Dashed line: mean  MW $A_V/N_\mathrm{H}$ ratio    
  for $R_V=3.1$ (Bohlin et al. 1978).  Dotted line: mean SMC
  $A_V/N_\mathrm{H}$ ratio   (Gordon 2003).
 {\bf (b)} 
  Dust-to-gas ratio, $A_V/N_\mathrm{H}$, versus metallicity for the same set of data.
  The labels indicate the regions of the diagram occupied by MW and SMC interstellar clouds.
     }
              \label{AV-NH}%
    \end{figure*}

\subsection{ The dust-to gas ratio   $A_V/N_\mathrm{H}$ }

In   Fig.  \ref{AV-NH}{\bf a} we plot $A_V$ versus the total hydrogen column
density for the same data set of Fig.  \ref{AV-NdFe}.
For DLA absorbers we assume that $N_\mathrm{H}=N(\ion{H}{i})$, i.e.
we neglect the molecular contribution. 
Two of our systems with reddening detections are not plotted
because they do not have a direct determination of $N(\ion{H}{i})$ (see Table 1).
This is also the case for the 
 \ion{Mg}{ii} absorbers  of York et al. (2006)
and   the \ion{Ca}{ii} absorbers of Wild et al. (2006)
discussed in Section 4.2.  

 The comparison with the interstellar data   shows that  
DLA systems do not follow the
typical Milky Way relation  of Bohlin et al. (1978),
shown as a dashed line in the figure. 
Most   systems lie below the MW trend and  
the addition of a molecular contribution to the
hydrogen column density would not change this conclusion.

The analysis of the data versus metallicity  reveals the
existence of a regular behaviour.
This can be seen
in   Fig.  \ref{AV-NH}{\bf b}, where we plot 
the dust-to-gas ratio $A_V/N_\mathrm{H}$ versus
metallicity [Zn/H]  for the same systems. 
In spite of the limited size of the DLA sample, the data suggest the
existence of a trend of increasing dust-to-gas ratio with
increasing metallicity. The trend nicely fits the regions of the diagram
occupied by MW and SMC interstellar clouds,
suggesting its general validity at low and high redshift.

The observed trend 
can be interpreted as a natural consequence of
the approximate constancy of $\langle s^\mathrm{Fe}_V \rangle$ discussed above.
In fact,
dividing both sides of Eq. (\ref{DeltaNorMetCol}) by $N_\mathrm{H}$ and inserting
relation (\ref{NFedust1}), we obtain
\begin{equation}
{ A_V \over N_\mathrm{H} }
= \langle s^\mathrm{Fe}_V \rangle ~
{  f_\mathrm{Fe} \over   (1-f_\mathrm{Zn}) } \,  { N_\mathrm{Zn} \over  N_\mathrm{H}    } \,
\left( {\mathrm{Zn} \over \mathrm{Fe} }\right)^{-1}_{a}
 ~.
\label{dust-to-gas}
\end{equation} 
If $\langle s^\mathrm{Fe}_V \rangle$ is constant, we expect 
from this expression a linear
trend between dust-to-gas ratio and metallicity (Zn/H),
modulated by variations
of the iron and  zinc depletions;  
 the intrinsic abundance ratio 
$\left( {\mathrm{Zn} \over \mathrm{Fe} }\right)_{a} $ is
not expected to show strong variations. 

If the trends of Fig. \ref{AV-NdFe} and \ref{AV-NH}  will be confirmed by a larger set of data, then 
it will be natural to  explain the validity of the Bohlin et al. (1978) MW relation,  $A_V/N_\mathrm{H} \sim$ constant, 
as the consequence of the constant metallicity level of MW clouds
in conjunction with a  constant $\langle s^\mathrm{Fe}_V \rangle$.

The deviations of the DLA dust-to-gas ratios 
from  the   Bohlin et al.'s value indicate that the MW  ratio should not be   applied   to obtain indirect
estimates of extinctions or $N_\mathrm{H}$ in extragalactic research. 
By taking into account the metallicity and the extinction curve of the absorbers
one can, however, obtain reasonable estimates. 
Since quasar absorbers are, in general, metal deficient, the typical    
$A_V/N_\mathrm{H}$ ratio  of the SMC (dotted line in Fig.  \ref{AV-NH})
is more appropriate than the MW ratio for indirect estimates of     extinctions or  
$N_\mathrm{H}$. 
The choice of an SMC dust-to-gas ratio is in line with the lack of 217.5\,nm bump
in most quasar absorbers  (York et al. 2006).

\section{Summary and conclusions}

We have implemented a technique for
measuring the reddening of   quasars    
using the photometric and spectroscopic SDSS database.   
For each quasar under investigation we build up a   control sample
to analyse the distribution
of the colors at a given redshift and magnitude.
Quasars showing absorption lines redwards of the Ly\,$\alpha$ emission
in their spectra
are rejected from the control sample. 
The median and the dispersion of the color distribution 
are used to estimate the reddening  and its  uncertainty.
To minimize systematic errors related to the presence
of quasar emissions in particular passbands,
the reddening  is measured in different color indices. 
%

We have applied this technique to the complete
sample of 13 SDSS quasars for which  
a single intervening DLA system with zinc absorption lines had been previously detected
in spectra  of high (or intermediate) resolution.
Most of the spectra of these quasars do not show evidence of
potential sources of reddening other than the DLA systems. 
We detect reddening at the 2\,$\sigma$ level in five of these quasars.
In each case the detection is confirmed in different color indices.
The comparison of the color excess measured in different pairs of bandpasses
is consistent with an origin of the reddening in the   DLA systems.
To our knowledge this is the most direct evidence 
  of DLA extinction up to $z \simeq 2$ obtained so far.

In order to  discriminate the extinction curve of the absorbers  
we have compared 
 the reddening in different color indices and investigated
the region of the SDSS spectra where the redshifted 217.5 nm MW-extinction bump
is expected to fall, if present.
An SMC-type extinction curve is generally consistent with the available data,
in agreement with previous studies of the extinction curves of 
\ion{Ca}{ii} and \ion{Mg}{ii} quasar absorbers  
(Wild \& Hewett 2005, York et al. 2006).
The only  possible 
exception is J0121+0027, for which we 
 cannot exclude that  the MW bump is present,
as  claimed by Wang et al. (2004).

After converting
the quasar reddening into $V$-band extinction in the rest frame of the DLA system,
$A_V$ (see Appendix A), 
we have investigated the relation between  $A_V$
and the dust-phase column density of iron
of the absorber, $\widehat{N}_\mathrm{Fe}$.
Our  measurements  and upper limits  
are consistent with a rise of $A_V$ with increasing $\widehat{N}_\mathrm{Fe}$,
but  the sample is too small to perform a correlation study.

By  comparing  these data 
with measurements of $A_V$ and $\widehat{N}_\mathrm{Fe}$
in Milky Way and SMC interstellar clouds,
we obtain the
main result of our work: the high-redshift data follow remarkably well
the trend of the  local interstellar data, which show a linear correlation  
$\log A_V$ - $\log \widehat{N}_\mathrm{Fe}$.
%
Only in one quasar, J2340$-$0053, do we find an excess of extinction,
possibly due to neutral gas
at a redshift different from that of the DLA system. 

Consistent results are found from the analysis of the  two quasars    
with    $(B-K)$ reddening detections  recently reported by Ellison et al. (2005):
one DLA system with \ion{Zn}{ii} lines follows
 the Milky-Way trend and another one, towards B0438$-$436, shows an excess
of extinction. In this case the excess is likely to originate in
dust of the quasar environment  (Ellison et al. 2005).

Finally, from the studies of Wild et al. (2006)
and York et al. (2006),
we find evidence that also the \ion{Ca}{ii} quasar absorbers at $z \sim 1$
and the \ion{Mg}{ii} absorbers 
with highest equivalent widths, statistically more
similar to DLA systems, follow the interstellar trend 
between extinction and dust-phase metal column density.

The existence of
a linear relation between $A_V$ and $\widehat{N}_\mathrm{Fe}$
shared by  high redshift DLA galaxies and by local clouds
of the Milky Way and the SMC
suggests that the mean extinction per atom of iron in the dust,
$\langle s^\mathrm{Fe}_V \rangle$,
is approximately constant in galaxies with different levels of metallicity
(from [Zn/H]$\simeq -1.4$ dex up to $\simeq +0.5$ dex)
and look-back times (up to $\sim 10$ billon years before the present). 
This result in turn suggests the existence of a mechanism which
tends to compensate variations of $\langle s^\mathrm{Fe}_V \rangle$
resulting from changes of the dust grain properties
in different interstellar environments.
We propose that when the grain size $a$
increases, the   density of the grain, $\varrho$, and the  
abundance by mass of iron in the dust, $X_\mathrm{Fe}$, 
decrease (and vice versa). 
These variations are expected to occur 
 when passing from a dust-destructive
to a dust-protected environment (and vice versa).

The existence of a well defined trend between metal column densities
and extinction in DLA systems at different redshifts
has important implications for quantifying the selection effect of quasar
obscuration (Fall \& Pei 1989).
The results of the present analysis lend quantitative support to
the estimates of DLA extinction performed   
in our previous study of the obscuration effect 
(Vladilo \& P\'eroux 2005).

Finally, from the analysis  of the dust-to-gas ratio
$A_V/N_\mathrm{H}$ in the DLAs of our sample,
we find significant deviations from the Milky-Way
 ratio $A_V/N_\mathrm{H} \approx 5.3 \times 10^{-22}$ mag cm$^{2}$ of Bohlin et al. (1978). 
The dust-to-gas ratio  appears to increase with the     metallicity
of the absorber, following a trend that fits the SMC and MW data points,
 but more data are necessary to confirm this result.
We argue that
the constancy of the dust-to-gas ratio in local interstellar clouds
is the result of  their constant level of metallicity in conjunction with
 the lack of significant variations of $\langle s^\mathrm{Fe}_V \rangle$.
  The MW dust-to-gas ratio should not be used in extragalactic studies 
 for indirect estimates of extinctions or \ion{H}{i} column densities.  

An important  outcome of the present investigation stems from the
comparison of the control samples before and after
the rejection of quasars
with  absorptions lines redwards of the Ly\,$\alpha$ emission.
After the rejection,   
the color distributions of the control samples experience   
 a small but systematic blue  shift 
 indicating that these absorption lines   do contribute to the 
reddening of quasars.  %
This result is in line with the 
 recent detection of reddening due to  
 \ion{Ca}{ii}  (Wild \& Hewett 2005, Wild et al. 2006) and
 \ion{Mg}{ii} (Khare et al. 2005; York et al. 2006)
absorption systems. 
Even in the absence of a direct measurement of the
\ion{H}{i} column densities of the  \ion{Ca}{ii}  and \ion{Mg}{ii} absorbers,
 the results of these  previous investigations provide indirect
 evidence that DLA systems 
should give a signal of reddening.  
The present investigation 
is in line with this expectation, even if the sample
is too small to draw general conclusions on the statistical
properties of the extinction in DLA systems.

At the present state of the observations it is hard to understand whether
or not there is a discrepancy between the reddening   detection
in  \ion{Ca}{ii},   \ion{Mg}{ii},   and DLA absorbers at $z \lsim 2$, on the one hand,
and the upper limit     found by Murphy \& Liske (2004) in DLA systems at $z \sim 2.8$,
on the other.
Future studies of extinction and metal column densities in DLA systems  
should be aimed at obtaining measurements at  higher redshifts
($z \gsim 2$),
at high values of extinctions ($A_V \gsim 1$ mag) and at
different values of $R_V$. 
With such measurements one should be able to understand if the relation
between extinction and dust-phase metal column density
is indeed universal, 
to probe the  redshift evolution of the dust in  cross-section selected galaxies,
to quantify with accuracy the effect of quasar obscuration,
and to probe the mechanism that makes  $\langle s^\mathrm{Fe}_\lambda \rangle$
relatively constant in interstellar clouds.

 \begin{acknowledgements} 
 The insightful comments of the referee
 have greatly improved the presentation of this work. 
    We thank Jason Prochaska for communicating data
    in advance of publication.
    The work of SAL is supported by the
RFBR grant No. 06-02-16489 and the LRSSP grant No. 9879.2006.2. 
    VPK acknowledges partial support from the US National Science 
Foundation grant AST-0206197 and the NASA/STScI grant GO 9441.
\end{acknowledgements}

\appendix
 

\section{Conversion of the quasar reddening into rest frame extinction}

Our goal is to  estimate the rest frame extinction, $A_V $, from
the measurement of the quasar reddening in the observer frame, $\Delta (y-x)$. 
We start from the definition of
the apparent magnitude of a quasar in a photometric band $x$,
\begin{equation}
m_x = m_{x,\circ} \, 
-2.5 \, \log \int_{\lambda_{x_1}}^{\lambda_{x_2}}
 \, F(\lambda) \, S_x(\lambda) \, d \lambda
\label{A1}
\end{equation}
where $m_{x,\circ}$ is the zero point of the photometric system;
$F(\lambda)$ is the monochromatic flux of the quasar
received out of the Earth's atmosphere
(erg s$^{-1}$ cm$^{-2}$ \AA$^{-1}$);
 $S_x(\lambda)$ is the response curve
of the passband $x$ (we include in this term the transmission 
of the Earth's atmosphere and the
response curve of the detector);
$\lambda_{x_1}$ and $\lambda_{x_2}$ are the lower and upper cutoff of the bandpass. 

The flux received at Earth from a quasar at redshift $z_e$ is
\begin{equation}
F(\lambda) =
{L({\lambda_e}) H^2_\circ (1+z_e)^{-3}
\over 
4 \pi c^2 Z^2_q(z_e)} = 
{ L({\lambda_e}) (1+z_e)^{-1} 
\over 
4 \pi D_L^2}\; ,
\label{A2}
\end{equation}
where 
the luminosity distance
$ D_L= c\,Z_q(z_e)\,(1+z_e)/H_\circ $
is defined through the function (Weinberg 1972)
$$
Z_q(z_e) = {1\over q_\circ^2(1+z_e)}
\left\{ q_\circ z_e  + (q_\circ -1) 
\left[ (1+2 q_\circ z_e)^{1 \over 2} - 1 \right] \right\}.
$$
Here
$L(\lambda)$ is the monochromatic luminosity emitted by the quasar
(erg s$^{-1}$  \AA$^{-1}$), and $\lambda_e=\lambda/(1+z_e)$.

\medskip\noindent
From this we have 
\begin{equation}
m_x = C_{x}(z_e) 
-2.5  \log \int_{\lambda_{x_1}}^{\lambda_{x_2}}
\!\! L({\lambda_e}) 
  \, S_x(\lambda) \, d \lambda
\label{A3}
\end{equation}
where the term $C_{x}(z_e)$ depends on the zero-point of the
photometric system, the quasar redshift and the cosmological parameters. 

We now consider the extinction due to 
an intervening interstellar medium $i$ that     lies
at the absorption redshift $z_\mathrm{abs}$ in the direction of the quasar. 
We call $S_i(\lambda)$ the transmission
of the medium as  a function of wavelength in the absorber rest frame
[$S_i(\lambda)=1$ for a transparent medium]. 
The reddened magnitude of the quasar is
\begin{equation}
m^\prime_x = C_{x}(z_e) 
-2.5  \log \int_{\lambda_{x_1}}^{\lambda_{x_2}} \!\!
  L({\lambda_e}) 
\,S_i({\lambda_a}) \, S_x(\lambda) \, d \lambda ~,
\label{A4}
\end{equation}
where $\lambda_a = \lambda/(1+z_\mathrm{abs})$. 

\medskip\noindent
By definition, the quasar extinction in the observer's frame is
$A^\prime_x \equiv m^\prime_x - m_x$.  From this we derive
the relation
\begin{equation}
A^\prime_x =  
-2.5  \log 
{
\int_{\lambda_{x_1}}^{\lambda_{x_2}}  L({\lambda_e}) 
\,S_i({\lambda_a}) \, S_x(\lambda) \, d \lambda 
\over
\int_{\lambda_{x_1}}^{\lambda_{x_2}}  L({\lambda_e}) 
 \, S_x(\lambda) \, d \lambda 
} ~.
\label{A5}
\end{equation}

We now express the transmission of medium in terms of the 
extinction 
in the absorber rest frame, $A_\lambda$ (magnitudes).
By introducing the normalized extinction curve
of the absorber, 
$\xi(\lambda) \equiv A_\lambda/A_V $,
we obtain
\begin{equation}
S_i(\lambda) = 10^{-0.4\, A_V\, \xi(\lambda) } ~.
\label{newA6}
\end{equation}

The reddening of the quasar  in the color index $(y-x)$ 
in the observer's frame is
$ \Delta (y-x) = A^\prime_y - A^\prime_x$. 
Combining the above expressions we derive  
\begin{eqnarray}
\Delta (y-x) =  
2.5  \log 
{
\int_{\lambda_{x_1}}^{\lambda_{x_2}}  
10^{-0.4\,A_V\, \xi(\lambda_a)} 
L({\lambda_e}) \, S_x(\lambda) \, d \lambda 
\over
\int_{\lambda_{x_1}}^{\lambda_{x_2}}  L({\lambda_e}) 
 \, S_x(\lambda) \, d \lambda 
} & &
\nonumber \\
-2.5  \log 
{
\int_{\lambda_{y_1}}^{\lambda_{y_2}}  
10^{-0.4\,A_V\, \xi(\lambda_a) } 
L({\lambda_e}) \, S_y(\lambda) \, d \lambda 
\over
\int_{\lambda_{y_1}}^{\lambda_{y_2}}  L({\lambda_e}) 
 \, S_y(\lambda) \, d \lambda 
}.\hspace{0.3cm} & &  
\label{A6}
\end{eqnarray}

The reddening  is independent of the zero point of
the photometric system and of the cosmological parameters.
The dependence of $\Delta(y-x)$ on
the redshift and the spectral distribution
of the quasar is  modest since the quasar
continuum varies smoothly in  
the bandpasses $[\lambda_{x_1},\lambda_{x_2}]$
and $[\lambda_{y_1},\lambda_{y_2}]$.
The spectral energy distribution can be approximated
with a power law 
$L(\lambda) \propto \lambda^{\alpha_\lambda}$
with spectral index $\alpha_\lambda = -(2+\alpha_\nu)$. 

To solve the above equation in $A_V$ we perform 
the change of variables
$$t(\lambda_a) \equiv - 0.4\,\ln(10)~\xi (\lambda_a)\, ,$$
and $$r \equiv - 0.4\,\ln(10)~\Delta (y-x) ~.$$
Then eq.(\ref{A6}) becomes
\begin{eqnarray}
e^r & = &
{
\int_{\lambda_{y_1}}^{\lambda_{y_2}}  
e^{A_V   t(\lambda_a) } 
L({\lambda_e}) \, S_y(\lambda) \, d \lambda 
\over
\int_{\lambda_{y_1}}^{\lambda_{y_2}}  L({\lambda_e}) 
 \, S_y(\lambda) \, d \lambda 
 }\,\times
\nonumber \\
& & \times \left[
{
\int_{\lambda_{x_1}}^{\lambda_{x_2}}  
e^{A_V   t(\lambda_a) } 
L({\lambda_e}) \, S_x(\lambda) \, d \lambda 
\over
\int_{\lambda_{x_1}}^{\lambda_{x_2}}  L({\lambda_e}) 
 \, S_x(\lambda) \, d \lambda 
} 
 \right]^{-1}\, .
\label{A7}
\end{eqnarray}

We now express the integrals in the form of sums, calculating the functions
in $n_x$ equispaced points $\lambda_i$ in the interval $[\lambda_{x_1},\lambda_{x_2}]$
and
$n_y$ equispaced points $\lambda_j$ in the interval $[\lambda_{y_1},\lambda_{y_2}]$.
We obtain 
\begin{equation}
e^r = 
{ \sum_{j=1}^{n_y} L_j S_j \, e^{A_V   t_j }
\over
\sum_{j=1}^{n_y}  L_j S_j } 
\times
\left(
{ \sum_{i=1}^{n_x} L_i S_i \, e^{A_V   t_i }
\over
\sum_{i=1}^{n_x} L_i S_i } 
\right)^{-1}\, ,
\label{A8}
\end{equation}
where
$S_i = S_x (\lambda_i)$, $S_j = S_y (\lambda_j)$,
$t_i = t[\lambda_i/(1+z_\mathrm{abs})]$, $L_i=L[\lambda_i/(1+z_e)]$. 

By introducing the weights 
$$ w_{x_i} = { L_i S_i  \over  \sum_{i=1}^{n_x} L_i S_i }  $$
and
$$ w_{y_j} = { L_j S_j \over  \sum_{j=1}^{n_y} L_j S_j }  $$
we obtain
\begin{equation}
e^r = 
{ \sum_{j=1}^{n_y} w_{y_j} \, e^{A_V   t_j }
} 
\times
\left(
{ \sum_{i=1}^{n_x} w_{x_i} \, e^{A_V   t_i }
  } 
\right)^{-1}\, ,
\label{NumSol}
\end{equation}
where
$\sum_{i=1}^{n_x} w_{x_i} = 1$ and $\sum_{j=1}^{n_y} w_{y_j} = 1$.

\medskip\noindent
The extinction $A_V$ can be calculated from this expression 
or estimated from its simplified version. An approximate  
value of $A_V$ can be derived in the following way.

\medskip\noindent
Let $m_x$  be the value of $i$ for which the weight $w_{x_i}$ is maximum,
i.e. 
$w_{m_x}=\max\{w_{x_i}\}_{i=1}^{n_x}$ and
$m_y$ the value of $j$  for which  $w_{y_j}$ is maximum, i.e.
$w_{m_y}=\max\{w_{y_j}\}_{j=1}^{n_y}$.
We take the terms 
$w_{m_x} \, e^{A_V   t_{m_x}}$
and 
$w_{m_y} \, e^{A_V   t_{m_y}}$
out of the sums and obtain
\begin{eqnarray}
e^r & = &
\left[
 w_{m_y} \, e^{A_V   t_{m_y}}
\sum_{j=1}^{n_y} { w_{y_j} \over w_{m_y}} \, e^{A_V  ( t_j - t_{m_y} )
} 
\right]
\times 
\nonumber \\
& & \times\left[
w_{m_x}\, e^{A_V  t_{m_x}}
\sum_{i=1}^{n_x} { w_{x_i} \over w_{m_x}} \, e^{A_V  ( t_i -  t_{m_x} )
  }
\right]^{-1}\, .
\label{A10}
\end{eqnarray}
If $t_i$ varies smoothly in the intervals
$[\lambda_{x_1},\lambda_{x_2}]$ and $[\lambda_{y_1},\lambda_{y_2}]$,
so that $A_V\, |t_i -  t_{m_x}| \ll 1$
and $A_V \, |t_j -  t_{m_y}| \ll 1$,
we can expand each exponent inside the sums
into a series $e^x \simeq 1+x$.
After some algebra, we obtain
\begin{eqnarray}
e^r & = & 
\left[
1 +  A_V  (\langle t \rangle_y - t_{m_y})  \right]
 \, e^{A_V    t_{m_y} }
\times 
\nonumber \\
& &
\times
\{
\left[
1 +  A_V  ( \langle t \rangle_x - t_{m_x})  \right]
 \, e^{A_V    t_{m_x} }
\}^{-1}\, ,
\label{A11}
\end{eqnarray}
where
$\langle t \rangle_x = \sum_{i=1}^{n_x}  w_{x_i} t_i$
and
$\langle t \rangle_y = \sum_{j=1}^{n_y}  w_{y_j} t_j$\, . 

\medskip\noindent
If $|\langle t \rangle_x - t_{m_x}| \ll 1$ and 
$|\langle t \rangle_y - t_{m_y} | \ll 1$,
then we obtain
$$r \simeq A_V (t_{m_y}-t_{m_x})~,$$
which is equivalent to 
\begin{equation}
A_V \simeq { \Delta(y-x) \over
 \xi({\lambda_{m_y}  \over  1+z_\mathrm{abs}}) - \xi({\lambda_{m_x} \over  1+z_\mathrm{abs}})}~.
\label{SimpleSol}
\end{equation}
For narrow bandpasses (\ref{SimpleSol}) gives the correct solution.

\section{The mean extinction per atom of iron
in the dust}

%

We consider a line of sight intersecting
dust embedded in a neutral region.
We call
$\mathcal{N}_d$  the number of dust grains per cm$^2$ along the line of sight,
$\sigma_g$  the geometrical cross section of a single grain  
and 
$Q_\lambda$ the extinction efficiency factor,
 i.e. the ratio between optical and geometrical cross section of the grain
 (Spitzer 1978).
The dust will consist of
different families of grains, 
each one with its own 
physical/chemical properties and, in particular,
 distribution of grain sizes $a$.
%
%
%
The extinction due to the $j$-th family     is
\begin{equation}
A_{\lambda}^{(j)}  = 1.086 ~  \mathcal{N}_d^{(j)} ~
 \langle \,  Q_\lambda ~ \sigma_g \,  \rangle_j
\label{Alj}
\end{equation} 
where
the term between angle brackets is an average  
calculated by integrating  $Q_\lambda^{(j)} ~ \sigma_g^{(j)}$
over the distribution of grain sizes $a$  (see e.g.   Pei 1992). 
The total extinction   is
\begin{equation}
A_\lambda  = 1.086 ~  \mathcal{N}_d \,
 \sum_j  w_j \,
 \langle \, Q_\lambda ~ \sigma_g \, \rangle_j \,\,  
\label{Alsum}
\end{equation} 
where the weight $w_j= \mathcal{N}_d^{(j)} /\mathcal{N}_d$
is the fraction of dust grains in each family.  

We now use iron as a tracer of the metals in the dust.
%
We call $V_g$ and $\varrho$   the volume
and the internal mass density  of a grain, respectively,
 and $X_\mathrm{Fe}$ the abundance by mass
of iron in the grain. 
 The dust-phase column density of iron atoms 
in   the $j$-th family  of grains is
\begin{equation}
\widehat{N}_\mathrm{Fe}^{(j)}
=   {    
\langle  \,V_g \, \varrho  ~ X_\mathrm{Fe}  \, \rangle_j
 \over  
 \mathrm{A_\mathrm{Fe}} \, m_\mathrm{H} }
 ~   \mathcal{N}_d^{(j)}
  ~,
\label{NFedj}
\end{equation}
where 
$\mathrm{A_\mathrm{Fe}}$ is the atomic mass of iron.
The average between angle brackets is   
calculated by integrating  
$V_g^{(j)} \, \varrho^{(j)}  ~ X_\mathrm{Fe}^{(j)} $
over the distribution of grain sizes. 
The total dust-phase column density of iron  
is
\begin{equation}
\widehat{N}_\mathrm{Fe}
=   {   
 \sum_j  w_j \,
 \langle  \,V_g \, \varrho  ~ X_\mathrm{Fe}  \, \rangle_j
 \over  \mathrm{A_\mathrm{Fe}} \, m_\mathrm{H} }  ~
\mathcal{N}_d
  ~.
\label{NFedsum}
\end{equation}

Combining (\ref{Alsum}) and (\ref{NFedsum}) we have
\begin{equation} 
A_\lambda 
 = ~ \langle s^\mathrm{Fe}_\lambda \rangle
 ~  \widehat{N}_\mathrm{Fe} 
\label{Alsum2}
\end{equation}
where  
\begin{equation}
\langle s^\mathrm{Fe}_\lambda \rangle
=   1.007 \! \times \! 10^{-22} ~
{   \sum_j w_j \,  \langle \, Q_\lambda ~ \sigma_g \, \rangle_j
\over
\sum_j w_j \,  \langle  \,V_g \, \varrho  ~ X_\mathrm{Fe}  \, \rangle_j
} 
\label{sFel}
\end{equation} %
is the mean optical cross section $\mathrm{(mag \, cm^2 )}$
per atom   of iron in the dust. 


\begin{thebibliography}{}
 
\bibitem[]{}
Adelman-McCarthy, J. K., Ag\"ueros, M. A., Allam, S. S., Anderson, K. S. J.
Anderson, S. F. et al. 2006, \apjs, 162, 38


\bibitem[]{} Akerman, C.J., Ellison, S.L., Pettini, M., Steidel, C.C.
2005, \aap, 440, 499

\bibitem[]{} Bohlin, R.C., Savage, B.D., \& Drake, J.F.
1978, \apj, 224, 132

\bibitem[]{} Boiss\'e, P., Le Brun, V., Bergeron, J., \& Deharveng, J.M.
1998, \aap, 333, 841

\bibitem[]{} Cardelli, J.A., Clayton, G.C., \& Mathis, J.S.
1988, \apj, 329, L33 (CCM)

\bibitem[]{} Cayrel, R., Depagne E., Spite, M., Hill, V., Spite, F.
et al. 2004, \aap, 416, 117

\bibitem[]{} Chen, Y.Q., Nissen, P.E., \& Zhao, G.
2004, \aap, 425, 697


\bibitem[]{} Draine, B.T. 2003, \araa, 41, 241


\bibitem[]{} Ellison, S.L., Hall, P.B., Lira, P. 2005, \aj, 130, 1345

\bibitem[]{} Ellison, S. L., Yan, L., Hook, I.M., Pettini, M., Wall, J.V., \& Shaver, P.
2001, \aap, 379, 393


\bibitem[]{} Fall, S.M. \& Pei, Y. 1989, \apj, 337, 7 

\bibitem[]{} Fitzpatrick, E.L. 1985, \apjs, 59, 77



\bibitem[]{} Foltz, C. B., Chaffee, F. H., Hewett, P. C., Weymann, R. J., Anderson, S. F.,
\& MacAlpine, G. M. 1989, \aj, 98, 1959


\bibitem[]{} Fukugita, M., Ichikawa, T., Gunn, J.E., Doi, M., Shimasaku, K., \& Schneider, D.P.
1996, \aj, 111, 1748

\bibitem[]{} Gordon, K.D., Clayton, G.C., Misselt, K.A., Landolt, A.U., \& Wolff, M.J.
2003, \apj, 594, 279

\bibitem[]{} Gratton, R., Carretta, E., Claudi, R., Lucatello, S., \& Barbieri, M.
2003, \aap, 404, 187

 
\bibitem[]{} Khare, P., Kulkarni, V.P., Lauroesch, J.T., York, D.G., Crotts, A.P.S. \&
Nakamura, O. 2004, \apj, 616, 86


\bibitem[]{} Khare, P., York, D.G.,  Vanden Berk, D., Kulkarni, V.P., Crotts, A.P.S.
et al. 2005, in Probing Galaxies through Quasar Absorption Lines, Proc. IAU Coll. No. 199,
eds. P.R. Williams et al., p. 427


\bibitem[]{} Kulkarni, V.P., Fall, S.M., Lauroesch, J.T., York, D.G., Welty, D.E.,
Khare, P., \& Truran, J.W. 2005, \apj, 618, 68

\bibitem[]{} Jenkins, E.B., Savage, B.D. \& Spitzer, L. 1986, \apj, 301, 355

\bibitem[]{} Junkkarinen, V. T., Cohen, R. D., Beaver, E. A.,
 Burbidge, E. M., Lyons, R. W., \&  Madejski, G. 2004, \apj, 614, 658
 
 
 \bibitem[]{} Lanzetta, K.M., Wolfe, A.M., \& Turnshek, D.A. 1995, \apj, 440, 435 
 

\bibitem[]{} Ledoux, C., Petitjean, P., \& Srianand, R. 2003, \mnras, 346, 209

\bibitem[]{} Lu, L., \& Wolfe, A. 1994, \aj, 108, 44

\bibitem[]{} Meiring, J., Kulkarni, V. P., Khare, P., Bechtold, J., York, D. G., Cui, J., 
Lauroesch, J. T., Crotts, A. P. S., \& Nakamura, O. 2006, \mnras, submitted

\bibitem[]{}
{Meurer, G.R.} 2004 in: A.N. Witt et al. (eds.),
\textit{Astrophysics of Dust, ASP Conf. Ser.} 309, 195

\bibitem[]{} Meyer, D.M., Lanzetta, K.M., \& Wolfe, A.
1995, \apj, 451, L13

\bibitem[]{} Mishenina, T.V., Kovtyukh, V.V., Soubiran, C., Travaglio, C., \& Busso, M.
2002, \aap, 396, 189

\bibitem[]{} Murphy, M.T., \& Liske, J. 2004, \mnras, 354, L31  

\bibitem[]{} Neckel, T., \& Klare, G. 1980, \aaps, 42, 251

\bibitem[]{} Nissen, P.E., Chen, Y.Q., Asplund, M., \& Pettini, M.
2004, \aap,  415, 993

\bibitem[]{} Pei, Y.C. 1992, \apj, 395, 130

\bibitem[]{} Pei, Y.C., Fall, S. M.,  \& Bechtold, J. 1991, \apj, 378, 6

\bibitem[]{} P\'eroux, C., Kulkarni, V.P., Meiring, J., Ferlet, R., Khare, P., 
Lauroesch, J.T., Vladilo, G., \&
York, D.G. 2006 \aap, in press (astro-ph/0601079)

\bibitem[]{} Petitjean, P., Srianand, R., \& Ledoux, C. 2000, \aap, 364, L26

\bibitem[]{} Petitjean, P., Srianand, R., \& Ledoux, C. 2002, \mnras, 332, 383

\bibitem[]{} Prochaska, J.X. \& Wolfe, A. 1999, \apj, 121, 369

\bibitem[]{} Rachford, B.L., Snow, T.P., Tumlinson, J., Shull, J.M. \& Blair, W.P.
2002, \apj, 577, 221

\bibitem[]{} Rao, S.M., Prochaska, J.X., Howk, J.C., \& Wolfe, A.,
2005, \aj, 129, 9



\bibitem[]{} Rao, S.M., Turnshek, D.A., \& Nestor, D.B. 2006, \apj, 636, 610


\bibitem[]{} Richards, G.T., Fan, X., Schneider, D.P., Vanden Berk, D.E., Strauss, M.A.
et al. 2001, \aj, 121, 2308

\bibitem[]{} Richards, G.T., et al. 2003, \aj, 126, 1131


\bibitem[]{} Ryan, S.G., Norris, J.E., \& Beers, T.C. 1996, \apj, 471, 254


\bibitem[]{} Savage, B.D., \& Sembach, K.R. 1996, \araa, 34, 279

\bibitem[]{} Schneider, D.P., Hall, P.B., Richards, G.T., Vanden Berk, D.E.,
Anderson, S.F. et al. 2005, \aj, 130, 367

\bibitem[]{} Snow, T.P., Rachford, B.L., Figoski, L.
2002, \apj, 573, 662

\bibitem[]{} 
{Spitzer, L.} 1978, 
\textit{Physical Processes in the Interstellar Medium}
(New York: Wiley Interscience)

\bibitem[2002]{V02a} Vladilo, G. 2002a, \apj, 569, 295

\bibitem[2002]{V02b} Vladilo, G. 2002b, \aap, 391, 407

\bibitem[]{} Vladilo, G. 2004, \aap, 421, 479 

\bibitem[]{}
Vladilo, G., Centuri\'on, M., Bonifacio, P., \& Howk, J.C. 2001, \apj, 557, 1007 

\bibitem[]{} Vladilo, G. \& P\'eroux, C. 2005, \aap, 444, 461
 
\bibitem[]{} Wang, J., Hall, P.B., Ge, J., Li, A., \& Schneider, D.P. 2004, \apj, 609, 589

\bibitem[]{} Weinberg, S. 1972, {\em Gravitation and Cosmology} (New York: Wiley)

\bibitem[]{}
Welty, D.E., Lauroesch, J.T., Blades J.C., Hobbs, L.M., \& York, D.G. 2001,
\apj, 554, L75

\bibitem[]{} Wild, V., \& Hewett, P.C. 2005, \mnras, 361, L30

\bibitem[]{} Wild, V., Hewett, P.C., \& Pettini, M. 2006, \mnras, in press (astro-ph/0512042)



\bibitem[]{} Wolfe, A. M., \& Briggs, F.H. 1981, \apj, 248, 460



\bibitem[]{} Wolfe, A. M., Gawiser, E., \& Prochaska, J. X.
2005, \araa, 43, 861




\bibitem[]{} Yip, C.W., Connolly, A.J., Vanden Berk, D.E., Ma, Z., Frieman, J.A. et al.
2004, \apj, 128, 2603




\bibitem[]{} York, D.G., Khare, P., Vanden Berk, D., Kulkarni, V.P., Crotts, A.P.S.
et al. 2006,  \mnras, in press (astro-ph/0601279)

 
\end{thebibliography}
\end{document}